\documentclass[a4paper,preprintnumbers,showpacs,twocolumn,superscriptaddress,nofootinbib,amsmath,amssymb]{revtex4-1}
\usepackage[dvips]{graphics}
\usepackage[hypertex]{hyperref}
\usepackage{color,hyperref}
\usepackage{times}
\usepackage{mathrsfs}
\hypersetup{
  colorlinks=true,
  linkcolor=blue,
  citecolor=magenta,
  filecolor=magenta,
  urlcolor=blue
}

\def\beq{\begin{equation}}
\def\eeq{\end{equation}}
\def\bear{\begin{eqnarray}}
\def\ear{\end{eqnarray}}

\usepackage{enumitem}
\usepackage{graphicx,subfigure}
\usepackage{dcolumn}
\usepackage{bm}
\usepackage{color}

\begin{document}

\title{Harmonic oscillations of neutral particles in the $\gamma$ metric}

\author{Bobir Toshmatov}
\email{bobir.toshmatov@nu.edu.kz}
\affiliation{Department of Physics, Nazarbayev University,
53 Kabanbay Batyr, 010000 Astana, Kazakhstan}
\affiliation{Ulugh Beg Astronomical Institute,
Astronomicheskaya 33, Tashkent 100052, Uzbekistan}

\author{Daniele Malafarina}
\email{daniele.malafarina@nu.edu.kz}
\affiliation{Department of Physics, Nazarbayev University,
53 Kabanbay Batyr, 010000 Astana, Kazakhstan}

\author{Naresh Dadhich}
\email{nkd@iucaa.in}
\affiliation{Inter-University Centre for Astronomy and Astrophysics,
Post Bag 4, Ganeshkhind, Pune 411 007, India}

\begin{abstract}

We consider a well-known static, axially symmetric, vacuum
solution of Einstein equations belonging to Weyl's class and
determine the fundamental frequencies of small harmonic
oscillations of test particles around stable circular orbits in
the equatorial plane. We discuss the radial profiles of
frequencies of the radial, latitudinal (vertical), and azimuthal
(Keplerian) harmonic oscillations relative to the comoving and
distant observers and compare with the corresponding ones in the
Schwarzschild and Kerr geometries. We show that there exist
latitudinal and radial frequencies of harmonic oscillations of
particles moving along the circular orbits for which it is
impossible to determine whether the central gravitating object is
described by the slowly rotating Kerr solution or by a slightly
deformed static space-time.

\end{abstract}

\maketitle

\section{Introduction}\label{sec-intr}

The Zipoy-Vorhees space-time, also known as $\gamma$ metric is an
asymptotically flat vacuum solution of Einstein's equations which
belongs to the Weyl class of static, axially symmetric
space-times~\cite{Zip,Vor}. The $\gamma$ metric is completely
characterized by two parameters, namely $M>0$, which is related to
the gravitational mass of the source and the deformation parameter
$\gamma>0$. In the case of $\gamma=1$, the Schwarzschild
space-time is recovered. On the other hand, the cases of
$\gamma>1$ and $\gamma<1$ correspond to oblate and prolate
spheroidal sources, respectively, thus, showing that the parameter
$\gamma$ can be considered as a deformation parameter and that for
$\gamma\neq 1$ the coordinates are not spherical.

From the no-hair theorem it is obvious that for $\gamma\neq 1$ the
line element does not describe a black hole. In fact, it can be
shown that the surface $r=2M$ corresponds to a genuine curvature
singularity for every value of $\gamma$ different from one
\cite{virb,Papadopoulos:PRD:1981}. The singular surface $r=2M$
must be regarded as an infinitely redshifted surface which
observationally may present features similar to the Schwarzschild
event horizon. However, for certain values of $\gamma$, the
spacetime presents some unique features that allow it to be
distinguished from the Schwarzschild solution.

For these reasons, the $\gamma$ metric can be considered as a
black hole  ``mimicker" and, being an exact solution of Einstein's
equations, constitutes an excellent candidate to study possible
astrophysical tests of black hole space-times.

The geometrical properties of the $\gamma$-metric have been
studied  in \cite{herrera1,herrera2,hernandez,Bonnor:GRG:1992}
while interior solutions have been found in
\cite{interior0,interior1,interior2}. The motion of test particles
and light rays in the $\gamma$ metric has been studied in
\cite{herrera1,disk,Boshkayev:PRD:2016,us1,Askar:arxiv:2019}.

The frequencies of quasiperiodic oscillations in the Schwarzschild
metric were studied for example in
\cite{Abramowicz:APSS:2005,Kolos:CQG:2015}, while for the Kerr
metric they have been studied in various physical scenarios, for
example in \cite{Kolos:EPJC:2017,Tursunov:PRD:2016,Torok:AA:2005}.
The application to astrophysics and the study of astrophysical
black holes was discussed in
\cite{Torok:AA:2005a,Bambi:JCAP:2012,Bambi:EPL:2016,Stuchlik:MNRAS:2015,Stuchlik:GRG:2009,Stuchlik:APJ:2016,Stuchlik:AA:2016}.

In the present article we study the frequencies of small harmonic
oscillations of test particles about stable circular orbits in the
$\gamma$ metric and compare them with the corresponding
frequencies in the Schwarzschild and Kerr space-times. We find
that differences appear at small radii (either approaching the
innermost stable circular orbit (ISCO) or approaching the
infinitely redshifted surface) and a combination of measurements
for epicyclic frequencies at small radii around compact objects
could be used in principle to determine the nature of its
geometry.

The paper is organized as follows: In Sec.~\ref{test-particles} we
recap the equations describing the motion of test particles in the
$\gamma$ metric and derive the values of $\gamma$ that separate
different behaviours. Sec.~\ref{frequencies} is devoted to the
study of small harmonic oscillations for test particles about the
circular geodesics. Finally in Sec. \ref{conc} the results are
summarized and put in the context of possible future astrophysical
observations of black holes. Throughout the paper we make use of
natural units setting $G=c=1$.

\section{Dynamics of test particles}\label{test-particles}

In Erez-Rosen coordinates \cite{ER} the $\gamma$ metric  is
represented by the line element
\bear\label{line-element}
ds^2=&&-f^\gamma
dt^2+f^{\gamma^2-\gamma}g^{1-\gamma^2}\left(\frac{dr^2}{f}
+r^2d\theta^2\right)+\nonumber\\&&+f^{1-\gamma}r^2\sin^2\theta
d\phi^2\ ,
\ear
where
\bear
&&f(r)=1-\frac{2M}{r}\ ,\\
&&g(r,\theta)=1-\frac{2M}{r}+\frac{M^2\sin^2\theta}{r^2}\ .\nonumber
\ear

From the asymptotic expansion of the gravitational potential it
is easy to see that the total mass of the source as measured by an
observer at infinity is $M_{\rm tot}=M\gamma$ \cite{hernandez}.
Also from evaluation of the Kretschmann scalar it is possible to
see that the surface $r=2M$ is a true curvature singularity for
all values of $\gamma \neq 1$ \cite{virb}. \footnote{However it is
also an infinite red-shift surface which means any signal
emanating from it would be infinitely red-shifted. It would
therefore be rather innocous for an external observer.} Therefore
the radial coordinate in the $\gamma$ space-time takes values
$r\in(2M,\infty)$. If we understand the singularity as the regime
at which the classical description fails, then we can interpret
the surface $r=2M$ as the boundary of an exotic compact object
that is intrinsically quantum-gravitational in nature. Our purpose
is to investigate the properties of test particles orbiting around
such an exotic compact object and determine whether they can in
principle be distinguished from the corresponding cases around a
black hole.

\subsection{Equations of motion}

Since the space-time under study does not depend explicitly on
time, the Hamiltonian plays the role of the total energy of the
system and governs the dynamics of neutral test particle. Such
Hamiltonian for test particles in curved space-time can written as
\bear\label{total-H}
H=\frac{1}{2}g^{\mu\nu}p_\mu p_\nu+\frac{1}{2}m^2\ ,
\ear
where $p^\mu$ is four-momentum that is defined as $p^\mu=m u^\mu$
with $m$ and $u^\mu$ being mass and four-velocity of the test
particle, respectively. From Noether's theorem, as the system does
not depend explicitly on the coordinates $t$ and $\phi$, we know
that the associated conjugate momenta, namely the energy $E$ and
angular momentum $L$ are conserved and \bear\label{conservations}
p_t=g_{tt}\frac{dt}{d\tau}=-E\ ,\qquad
p_\phi=g_{\phi\phi}\frac{d\phi}{d\tau}=L\ . \ear From
Eq.~(\ref{conservations}) one can find the $t$ and $\phi$
components of the four-velocity of the test particle as \bear
\frac{dt}{d\tau}=-g^{tt}E\ ,\qquad
\frac{d\phi}{d\tau}=g^{\phi\phi}L\ . \ear and thus we obtain
\bear\label{total-H1}
H=\frac{1}{2}g^{rr}p_r^2+\frac{1}{2}g^{\theta\theta}p_\theta^2+H_{t\phi}\
, \ear with \bear\label{Hrphi}
H_{t\phi}(r,\theta)=\frac{1}{2}\left(g^{tt}E^2+g^{\phi\phi}L^2+m^2\right)\
. \ear The normalization condition $u_\mu u^\mu=-1$ leads to
$H=0$. Then, from Eq.~(\ref{total-H1}) one finds that
\bear\label{eq-motionr} g_{rr}\left(\frac{dr}{d\tau}\right)^2+
g_{\theta\theta}\left(\frac{d\theta}{d\tau}\right)^2
=-\frac{2H_{t\phi}(r,\theta)}{m^2}. \ear If one considers a
particle moving on a plane with $\theta_0={\rm const.}$, as is the
case for particles in accretion disks that are confined near the
equatorial plane $\theta=\pi/2$, then the equation of
motion~(\ref{eq-motionr}) takes the form \bear
g_{rr}\left(\frac{dr}{d\tau}\right)^2=R(r)
\equiv-\frac{2H_{t\phi}(r,\theta_0)}{m^2}\ , \ear where radial
function $R(r)$ can be written as \bear \label{radial-function}
R(r)=\left(1-\frac{2M}{r}\right)^{-\gamma}
\left(E^2-V_{eff}(r)\right), \ear with \bear\label{eff-potential}
V_{eff}(r)=\frac{L^2}{r^2}\left(1-\frac{2M}{r}\right)
^{2\gamma-1}+\left(1-\frac{2M}{r}\right)^\gamma. \ear If, on the
other hand, the particle is moving along a circular orbit on
$r=r_0$ with $\theta\neq {\rm const.}$, then the equation of
motion~(\ref{eq-motionr}) takes the form \bear
g_{\theta\theta}\left(\frac{d\theta}{d\tau}\right)^2=
\Theta(\theta)\equiv-\frac{2H_{t\phi}(r_0,\theta)}{m^2}\ , \ear By
combining the two conditions above, for a particle on circular
orbit $r=r_0$ in the plane $\theta=\theta_0$, we obtain
\bear\label{circ-r} R(r_0)=0\ ,\qquad \partial_rR(r)|_{r_0}=0\ ,
\ear (or $V_{eff}(r_0)=E^2$ and $\partial_r V_{eff}(r_0)=0$) and
\bear\label{circ-theta} \Theta(\theta_0)=0\ ,\qquad
\partial_\theta\Theta(\theta)|_{\theta_0}=0\ , \ear

\subsection{Circular orbits}

In the following we shall focus on the radii of characteristic
circular orbits in the equatorial plane of the $\gamma$
space-time, since these are the most relevant orbits for
astrophysical purposes, as they describe the motion of particles
of gas in accretion disks around compact objects.

By solving Eqs.~(\ref{circ-r}), simultaneously, one finds the
specific energy $E=E/m$ and the specific angular momentum $L=L/m$
of test particles moving along circular orbits as
\bear
&&E=-\frac{g_{tt}}{\sqrt{-(g_{tt}+g_{\phi\phi}\Omega^2)}},\label{ener}\\
&&L=\frac{g_{\phi\phi}\Omega}{\sqrt{- (g_{tt}+g_{\phi\phi}\Omega^2)}},\label{angul}
\ear
Here $\Omega=d\phi/dt$ is the angular velocity of the test
particle as measured by distant observers and it is given by
\bear\label{ang-vel}
\Omega=\pm\sqrt{-\frac{g_{tt,r}}{g_{\phi\phi,r}}}\ . \ear Thus, we
rewrite expressions~(\ref{ener}), (\ref{angul}) and
(\ref{ang-vel}) in terms of the $\gamma$-metric as \bear
&&E=\left(1-\frac{2 M}{r}\right)^{\gamma/2} \sqrt{\frac{r-\gamma
 M-M}{r-2\gamma M-M}}\ ,\label{ener1}\\
&&L=\pm r\left(1-\frac{2 M}{r}\right)^{(1-\gamma)/2} \sqrt{\frac{M\gamma}{r-M-2M\gamma}}\ ,\label{angul1}\\
&&\Omega=\pm\left(1-\frac{2
M}{r}\right)^{\gamma-1/2}\frac{1}{r}\sqrt{\frac{M\gamma}{r-M-M\gamma}}\
.\label{ang-vel1} \ear From the above expressions one can see that
the regions of the $\gamma$ space-time where test particles can
have circular orbits are given by: \bear \label{circ-cond}
 (i)&&\ \quad r>M(2\gamma+1) \quad {\rm if} \quad \gamma\geq\frac{1}{2}\ , \nonumber\\
 (ii)&&\ \quad r>2M \quad {\rm if} \quad 0<\gamma<\frac{1}{2}.
\ear Moreover, from the fact that at the light ring (photon's
capture orbits) the specific energy~(\ref{ener1}) diverges because
photon's mass is zero and the relevant parameter is the ratio,
$L/E$, one can easily find the location of photon capture orbit as
\bear\label{photonsphere} r_{ps}=M(2\gamma+1)\ , \quad {\rm with}
\quad \gamma\geq\frac{1}{2}\ , \ear while for $\gamma<1/2$ no
photon capture orbit is present. By comparing~(\ref{circ-cond})
with (\ref{photonsphere}) one can say that innermost positions of
circular orbits of test particles are limited by the circular
geodesics of massless particles. That is, photon circular orbit
defines the existence threshold, $r>r_{ps}$, for timelike circular
orbits. Further note that $r_{ps}$ is always preceded by unstable
circular orbits.

Circular orbits with $r>r_{ps}$ are therefore unstable and slight
departures from circularity leads to unbound motion. Unbound
orbits are separated from the bound orbits by a critical geodesic
called the marginally bound circular orbit $r_{mb}$ that is found
from the zero binding energy $E_{bind}\equiv
E(\infty)-E(r_{mb})=0$~\cite{Bardeen:APJ:1972,Hod:PRD:2013}. Since
the $\gamma$ space-time is asymptotically flat, $E(\infty)=1$
therefore, $E(r_{mb})=1$ gives \bear\label{mbo}
r_{mb}-M-2M\gamma-(r_{mb}-M-M\gamma)\left(1-\frac{2 M}{r_{mb}}\right)^{\gamma}=0\ ,\nonumber\\
\ear A particle with energy $E > 1$ that slightly departs from
circular  orbit has unbound motion. This means that the circular
orbit is unstable in such a way that with an infinitesimal small
outward perturbation, the particle will escape to infinity on an
asymptotically hyperbolic trajectory. On the other hand, for
$E<1$, perturbing a particle on an unstable circular orbit would
lead to bound motion. In the case of $\gamma=1$, by solving
Eq.~(\ref{mbo}), one obtains the marginally bound circular orbit
in the Schwarzschild space-time, $r_{mb}=4M$.

\begin{figure}[th]
\centering
\includegraphics[width=0.48\textwidth]{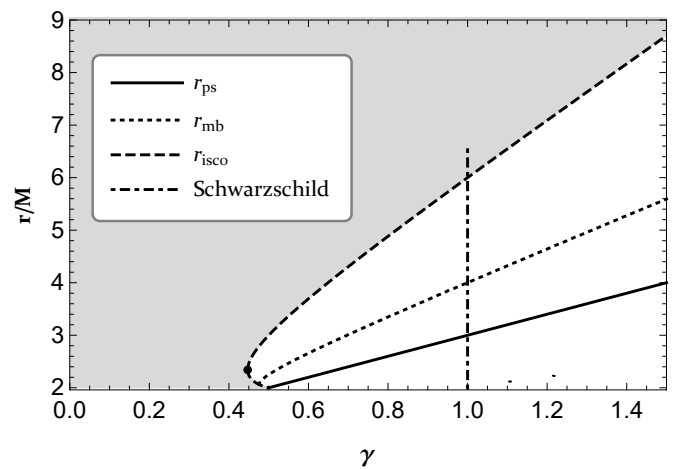}
\caption{\label{fig-orbits} Dependence of radii of characteristic
circular orbits: light ring ($r_{ps}$-black, solid), marginally
bound orbit ($r_{mb}$-black, dotted), and ISCO ($r_{isco}$-black,
dashed) in the $\gamma$ metric as functions of $\gamma$. Here the
vertical dot-dashed line corresponds to the values for the
Schwarzschild space-time with $\gamma=1$. The curve for
$r_{isco,-}$ for $\gamma>1/2$ is not shown since marginally stable
circular orbits are not allowed in this case. The gray region
corresponds to the range of values of $r$ and $\gamma$ where
stable circular orbits are allowed.}
\end{figure}

Now we find the marginally stable circular orbits, also called
innermost stable circular orbits (ISCO). All the stable circular
orbits of test particles (with radius $r_{st}$) satisfy the
condition ( $V_{eff}''(r_{st})\geq0$) or \bear\label{stable}
&&R''(r)|_{r_{st}}=\\
&&\frac{2M\gamma\left[r_{st}^2-2M(1+3\gamma) r_{st}+2M^2(1+
2\gamma)(1+\gamma)\right]}{r_{st}^2(r_{st}-2M)(M+2M
\gamma-r_{st})}\leq0,\nonumber \ear with the equality holding for
the smallest allowed value for stable circular orbits, namely the
ISCO. By solving the equation $R''(r)=0$ (or $V''(r)=0$), one
finds two roots for the ISCO given by
\bear\label{isco}
r_{isco,\pm}=M\left(1+3\gamma\pm\sqrt{5\gamma^2-1}\right)\ .
\ear
Once again, for $\gamma=1$ we retrieve $r_{isco,+}=6M$ which is
the value of the ISCO for the Schwarzschild geometry. Here it is
easy to notice that for $\gamma=1/\sqrt{5}$, the two values of the
ISCO coincide. Moreover, for $\gamma\geq 1/\sqrt{5}$ the value of
$r_{isco,+}$ is always greater than the bounds imposed by
conditions~(\ref{circ-cond}) and therefore there is always at
least one marginally stable circular orbit. However, $r_{isco,-}$
does not satisfy the bounds imposed by
conditions~(\ref{circ-cond}) for  values $\gamma\in(1/2,\infty)$.
Although it is natural to expect that stable circular orbits will
be allowed at great distances and cease to exist at a certain
distance from the source, the above discussion shows that in the
range $\gamma\in[1/\sqrt{5},1/2]$ there exist a second range of
stable circular orbits closer to center (see
Fig.~\ref{fig-orbits}). We can understand better the reason for
this behaviour by analyzing the effective potential in
eq.~(\ref{eff-potential}) term by term. It is easy to see that at
small radii the term proportional to $L^2/r^2$ will dominate over
the other terms. Then, as $\gamma<1/2$ the term
$(1-2M/r)^{2\gamma-1}$ will go at the denominator causing the
change in the behaviour of the effective potential. This new
region where stable circular orbits are allowed extends from
$r=2M$ until a finite distance, determined by the second root of
the ISCO equation. However, for $\gamma=1/\sqrt{5}$ the two values
of the ISCO radii coincide and for $\gamma<1/\sqrt{5}$ stable
circular orbits are allowed at any distance from the source,
similarly to the Newtonian case.

\begin{figure*}[th]
\centering
\includegraphics[width=0.48\textwidth]{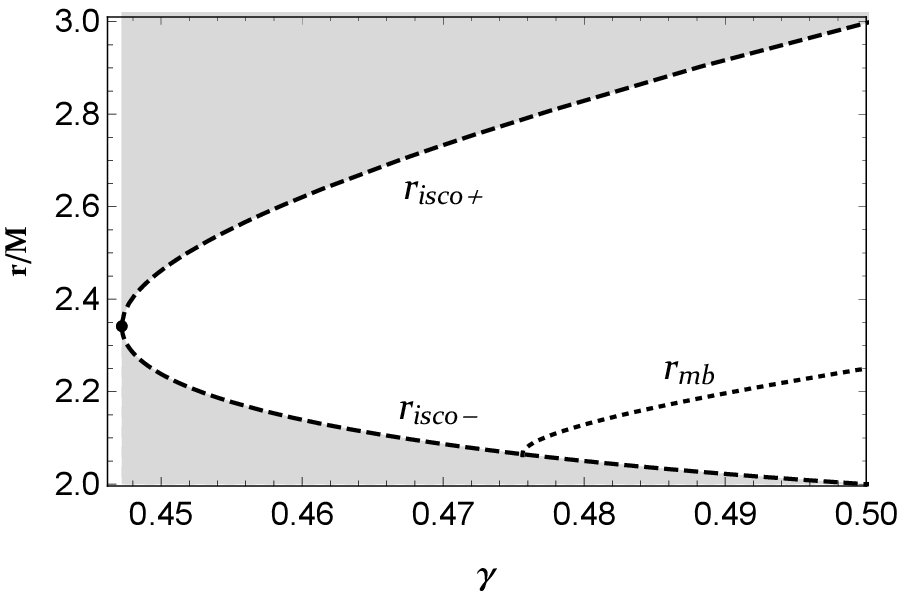}
\includegraphics[width=0.48\textwidth]{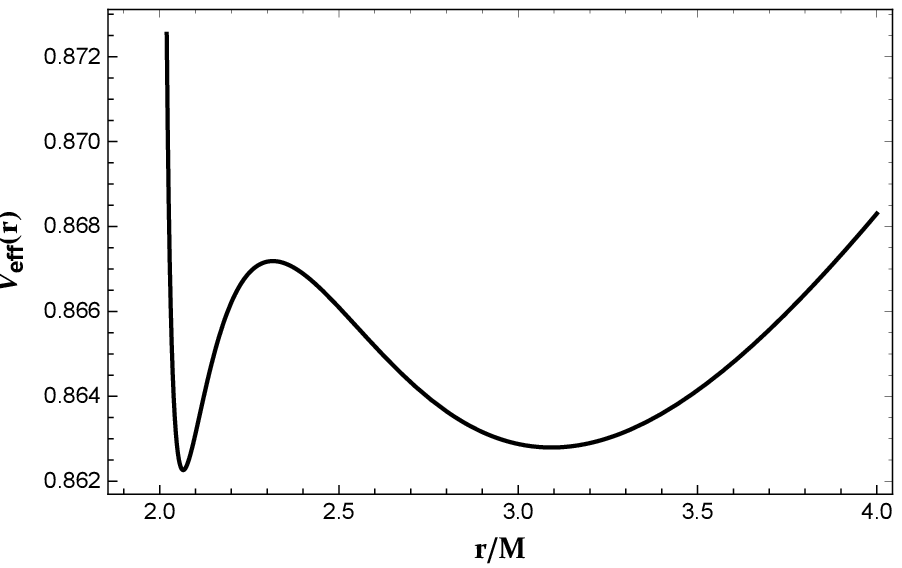}
\caption{\label{fig-potential} Left panel: Boundary of the ISCO
radii for $\gamma\in[1/\sqrt{5},1/2]$ from Fig.~\ref{fig-orbits}.
Here the gray shaded region represents the range of radii for
which stable circular orbits can exist as function of $\gamma$,
while the white region corresponds to the range for which no
stable circular orbits exist. The value of $r_{mb}$ separates the
regime where unstable orbits are bound from that were they are
unbound. Right panel: Radial profile of the effective potential
$V_{eff}$ in eq.~(\ref{eff-potential}) for
$\gamma\in(1/\sqrt{5},1/2)$. There are two minima of $V_{eff}$
corresponding to the stable circular orbits (where
$V_{eff}''(r_{st})\geq0$), and one maximum corresponding to the
unstable circular orbit (where $V_{eff}''(r_{unst})<0$).}
\end{figure*}

In order to extract more information on the region of the stable
circular orbits we can plot the effective potential $V_{eff}(r)$
in eq.~(\ref{eff-potential}) for different values of
$\gamma\in(0,1/2)$ (see Fig.~\ref{fig-potential} for the case of
$\gamma\in(1/\sqrt{5},1/2)$).
Depending on the value of $\gamma$ the effective potential
$V_{eff}$ can have up to two minima and one maximum. From the
condition~(\ref{stable}) one can easily realize that maxima
(minima) of the radial function (effective potential) correspond
to stable circular orbits. Then, the outer edge of the inner
stable circular orbits corresponds to $r_{isco-}$, while the inner
edge of the outer stable circular orbits corresponds to
$r_{isco+}$, i.e. the two solutions of the ISCO equation,
respectively. The maximum of the effective potential corresponds
to the unstable circular orbits. The effective potential exhibits
this behaviour with two minima and and one maximum precisely for
$1/\sqrt{5}<\gamma<1/2$ -- see right panel of
Fig.~\ref{fig-potential}. This shows that in two disjoint regions
$r_{st}\in(2M, r_{isco-}]$ and $r_{st}\in[r_{isco+},\infty)$ the
circular orbits are stable. However, in the region between these,
i.e. $r_{unst}\in(r_{isco-},r_{isco+})$, no stable circular orbits
can exist. For $\gamma>1/2$, the behaviour of the radial
function~(\ref{radial-function}) becomes similar to the
Schwarzschild case, where stable circular orbits are possible only
for $r_{st}\in[r_{isco+},\infty)$. At $\gamma=\sqrt{5}$ the outer
and inner minima merge and for $\gamma\leq1/\sqrt{5}$ the
behaviour of the radial function resembles the Newtonian case and
stable circular orbits are allowed everywhere. This suggests that
the term responsible for the angular momentum interaction with the
mass (the term $L^2M/r^3$ in the Schwarzschild case) behaves
qualitatively similar to the Schwarzschild case for $\gamma>1/2$,
namely it is attractive causing the stable circular orbits to
cease to exist at a certain radius, while it behaves in the
opposite way for $\gamma\leq 1/2$. The interaction between angular
momentum and mass turns from attractive to repulsive as $\gamma$
goes from $\gamma>1/2$ to $\gamma<1/2$. This can be clearly seen
by studying the term
\bear
W(r)=\frac{L^2}{r^2}\left(1-\frac{2M}{r}\right)^{2\gamma-1},
\ear
in the effective potential~(\ref{eff-potential}). It is
immediately seen for $\gamma=1/2$ we have $W=L^2/r^2$ and
therefore the relativistic correction to the Newtonian behaviour
vanishes. Then from the fact that $W'(r)=dW/dr$ near $r=2M$
changes sign at $\gamma=1/2$ we see that $W$ changes from
increasing (attractive) for $\gamma>1/2$ to decreasing (repulsive)
for $\gamma<1/2$ in the vicinity of the singularity.

Thus, taking into account the conditions~(\ref{circ-cond}), we can
conclude that the stable circular orbits can exist in different
regions as follows:
\bear
 &&\ \quad r\in[r_{isco+},+\infty) \quad {\rm for} \quad \gamma\in[1/2,+\infty)\ , \nonumber\\
 &&\ \quad r\in(2M,r_{isco-}]\cup[r_{isco+},+\infty) \quad {\rm for} \quad \gamma\in[1/\sqrt{5},1/2), \nonumber\\
  &&\ \quad r\in(2M,+\infty) \quad {\rm for} \quad \gamma\in(0,1/\sqrt{5})\ . \nonumber
\ear
It is interesting to notice that through the behaviour of the
motion of test particles we can gain some insight on the nature of
the curvature singularity in the space-time. In the case of
Schwarzschild, the singularity located at $r=0$ is attractive and
particles are crushed in it by diverging tidal forces. Similarly,
in the case of the $\gamma$ metric we can see that the same
behaviour occurs at $r=2M$ only for large values of $\gamma$. This
can be seen for example from the Kretschmann scalar $K$, which
behaves like $1/(r-2M)^6$ for $\gamma=2$. Therefore if we
characterize the strength of the singularity by the exponent $s$
for which $K\simeq 1/(r-2M)^s$, we see that $s>6$ for $\gamma>2$,
$s>4$ for $\gamma>1+\sqrt{5}$, and $s<2$ for $\gamma<1$,
suggesting that for prolate sources the singularity is weaker than
in the Schwarzschild case. For $\gamma>1/2$ the singularity is
'attractive' as can be seen from the fact that the interaction
between mass and angular momentum vanishes at $r=2M$. On the other
hand, when $\gamma<1/2$, the interaction term between the mass and
the angular momentum of the test particle is repulsive and blows
up at $r=2M$, as in the case of the monopole-quadrupole solution
of the Weyl class, which also describes small deviations from
spherical symmetry~\cite{Herrera:FPL:2005}. Therefore, in this
case, the singularity is ``repulsive" and test particles will be
ejected rather than crushed.

\section{Harmonic oscillations of neutral test particle} \label{frequencies}

In this section we will study the motion of test particles that
slightly depart from circular orbit. As we mentioned in the
previous section, stable circular orbits for test particles in the
$\gamma$ space-time are located at a radius $r_0$ and latitudinal
angle $\theta_0=\pi/2$ corresponding to the minimum of the
function $H_{t\phi}$ (or $R(r)$ and $\Theta(\theta)$,
respectively) given in Eq.~(\ref{Hrphi}). If a test particle
deviates slightly from the stable circular orbit it will start to
oscillate around its equilibrium value performing an epicyclic
motion. This oscillating motion is governed by the epicyclic
frequencies.

\subsection{Epicyclic frequencies}

For the sake of clarity we shall restrict the analysis to the
linear regime, and consider separately the case of purely radial,
i.e. $r=r_0+\delta r$, $\delta\theta=0$, and purely vertical,i.e.
$\theta=\theta_0+\delta\theta$, $\delta r=0$, epicyclic
oscillations about circular orbits in the equatorial plane
$\theta_0=\pi/2$. If we write the Taylor expansion of the
functions $R(r)$ and $\Theta(\theta)$ in powers of $\delta r$ and
$\delta\theta$, respectively as \bear
&&R(r)=R(r_0)+\partial_r R(r)|_{r_0}\delta r+\frac{1}{2}\partial_r^2
 R(r)|_{r_0}\delta r^2+...\ ,\\
&&\Theta(\theta)=\Theta(\theta_0)+\partial_\theta
\Theta(\theta)|_{\theta_0}\delta\theta+\frac{1}{2}\partial_\theta^2 \Theta(\theta)|_{\theta_0}\delta\theta^2+...\ ,
\ear
and apply the circularity conditions~(\ref{circ-r}) and~(\ref{circ-theta}), we arrive at
\bear
&&g_{rr}\delta \dot{r}^2=\frac{1}{2}\partial_r^2 R(r)|_{r_0}\delta r^2\ ,\\
&&g_{\theta\theta}\delta \dot{\theta}^2=\frac{1}{2}\partial_\theta^2 \Theta(\theta)|_{\theta_0}\delta \theta^2\ .
\ear
Considering that the total energy of the orbit is conserved, one
finds that
\bear
&&\delta \dot{r}\left[g_{rr}\delta\ddot{ r}-\frac{1}{2}\partial_r^2 R(r)|_{r_0}\delta r\right]=0\ ,\label{delta-r}\\
&&\delta\dot{\theta}\left[g_{\theta\theta}\delta \ddot{\theta}-
\frac{1}{2}\partial_\theta^2\Theta(\theta)|_{\theta_0}\delta \theta\right]=0\ ,\label{delta-theta}
\ear
Where $\delta\dot{r}=0$ and $\delta\dot{\theta}=0$ are the
trivial solutions corresponding to no oscillations. It is
immediately apparent that the parts within the square brackets of
the above equations are in the form of harmonic oscillators
as~\cite{Abramowicz:APSS:2005,Toshmatov:PRD:2017}
\bear
&&\delta\ddot{r}+\omega_r^2\delta r=0\ ,\\
&&\delta\ddot{\theta}+\omega_\theta^2\delta\theta=0\ ,
\ear
with
\bear
&&\omega_r^2=-\frac{\partial_r^2 R(r)|_{r_0}}{2g_{rr}}=
\frac{\partial_r^2H_{t\phi}|_{r_0}}{g_{rr}}\ ,\label{omega-r}\\
&&\omega_\theta^2=-\frac{\partial_\theta^2 \Theta(\theta)|_{\theta_0}}{
2g_{\theta\theta}}= \frac{\partial_\theta^2H_{t\phi}|_{\theta_0}}{g_{\theta\theta}}\ ,\label{omega-theta}
\ear
where the specific energy and angular momentum of the test
particles on circular orbits are given by Eq.~(\ref{ener1}) and
Eq.~(\ref{angul1}), respectively. Therefore we can obtain the
explicit forms of the frequencies as
\begin{widetext}
\bear
&&\omega_r^2=\left(1-\frac{M}{r}\right)^{2\gamma^2-2}
\left(1-\frac{2M}{r}\right)^{-\gamma^2+\gamma-1}\frac{M\gamma
\left[r^2-2(3\gamma+1)Mr+2(\gamma +1)(2\gamma +1) M^2\right]}{r^4(r-M-2M\gamma)}\ ,\label{omega-r1}\\
&&\omega_\theta^2=\left(1-\frac{M}{r}\right)^{2\gamma^2-2}
\left(1-\frac{2M}{r}\right)^{-\gamma^2+\gamma} \frac{M\gamma
}{r^2(r-M-2M\gamma)}\ ,\label{omega-theta1}
\ear
\end{widetext}
By inspecting expression (\ref{omega-r1}) one can notice that
zeros of expression in the square bracket correspond to the radii
of the ISCO as given in Eq.~(\ref{isco}). Thus, we see that radial
epicyclic frequency vanishes at the ISCO, $\omega_r(r_{isco})=0$.
This is reasonable, considering the fact that the ISCO is a
marginally stable orbit and that below ISCO, at least for values
of $\gamma>1/2$, the particles start to fall towards the central
object, and no radial oscillations occur.

Moreover, there is another very important angular frequency of
circular epicyclic motion of the particle, namely the azimuthal
(or Keplerian) frequency of the circular motion in the equatorial
plane, defined by the relation \bear\label{omega-phi}
\omega_\phi^2=\dot{\phi}^2=\left(1-\frac{2M}{r}\right)^{\gamma-1}
\frac{M\gamma}{r(r-M-2M\gamma)}\ . \ear

Note that the frequencies $\omega_r$, $\omega_\theta$ and
$\omega_\phi$ are measured with respect to the proper time of a
comoving observer. To get the observed frequencies by a observer
at infinity, one needs to divide them by the square of the
redshift factor $u^t$ as \bear
\Omega_{r(\theta,\phi)}^2=\frac{\omega_{r(\theta,\phi)}^2}{(u^t)^2}
\ , \ear where squared redshift factor for the $\gamma$ metric is
given by \bear
(u^t)^2=\left(1-\frac{2M}{r}\right)^{-\gamma}\frac{r-M-M\gamma
}{r-M-2M\gamma}\ . \ear Thus,
frequencies~(\ref{omega-r1}),~(\ref{omega-theta1}) and
(\ref{omega-phi}) take the following form when they are measured
by observers at infinity:
\begin{widetext}
\bear
&&\Omega_r^2=\left(1-\frac{M}{r}\right)^{2\gamma^2-2}
\left(1-\frac{2M}{r}\right)^{-(\gamma-1)^2}\frac{M\gamma
\left[r^2-2(3\gamma+1)Mr+2(\gamma +1)(2\gamma +1) M^2\right]
}{r^4(r-M-M\gamma)}\ ,\label{omega-r2}\\
&&\Omega_\theta^2=\left(1-\frac{M}{r}\right)^{2\gamma^2-2}
\left(1-\frac{2M}{r}\right)^{-\gamma^2+2\gamma}\frac{M\gamma
}{r^2(r-M-M\gamma)}\ ,\label{omega-theta2}\\
&&\Omega_\phi^2=\left(1-\frac{2M}{r}\right)^{2\gamma-1}
\frac{M\gamma}{r^2(r-M-M\gamma)}\ .\label{omega-phi2}
\ear
\end{widetext}
\begin{figure*}[th]
\centering
\includegraphics[width=0.33\textwidth]{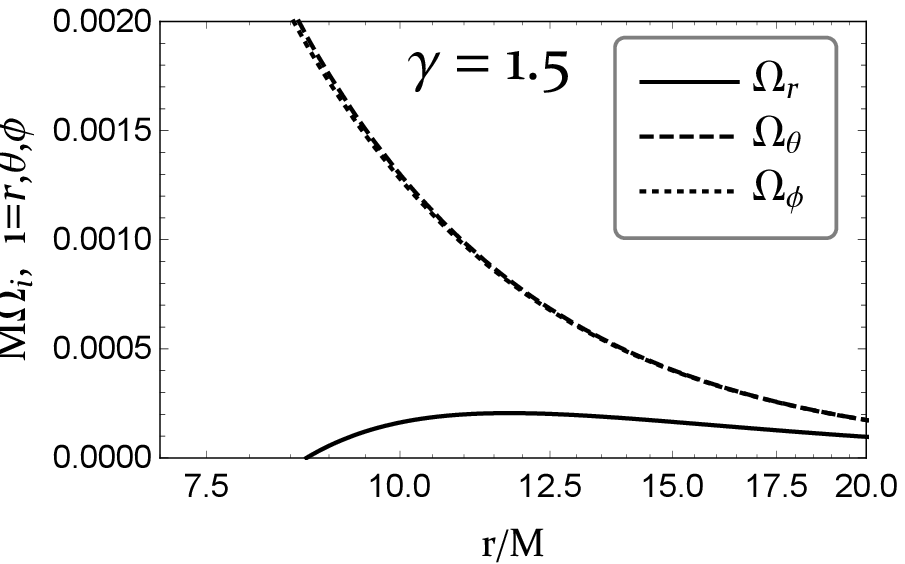}
\includegraphics[width=0.33\textwidth]{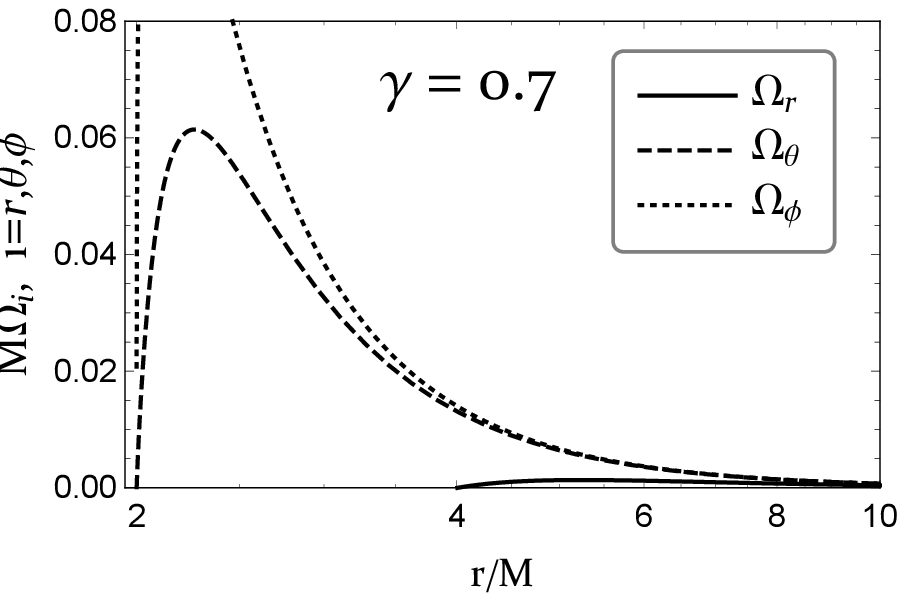}
\includegraphics[width=0.33\textwidth]{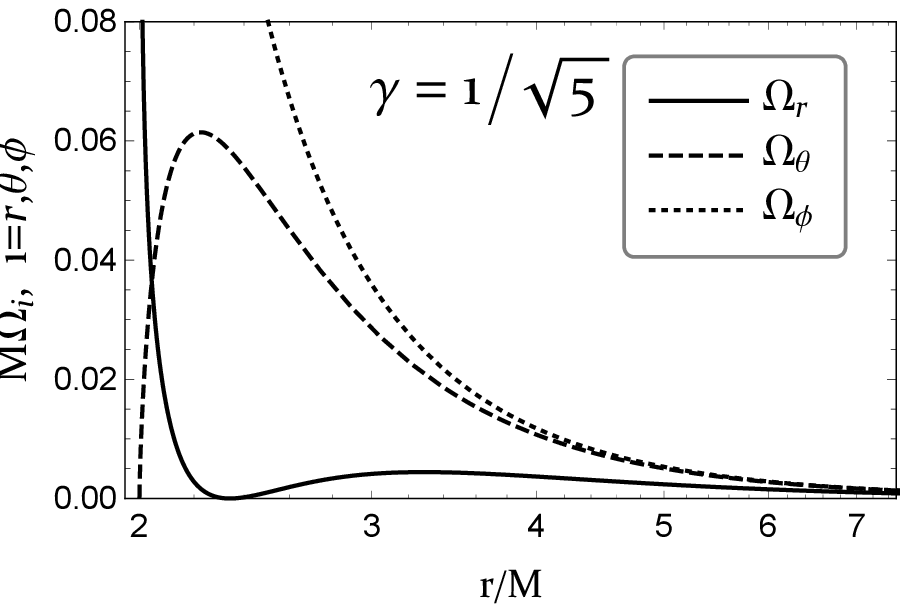}
\includegraphics[width=0.33\textwidth]{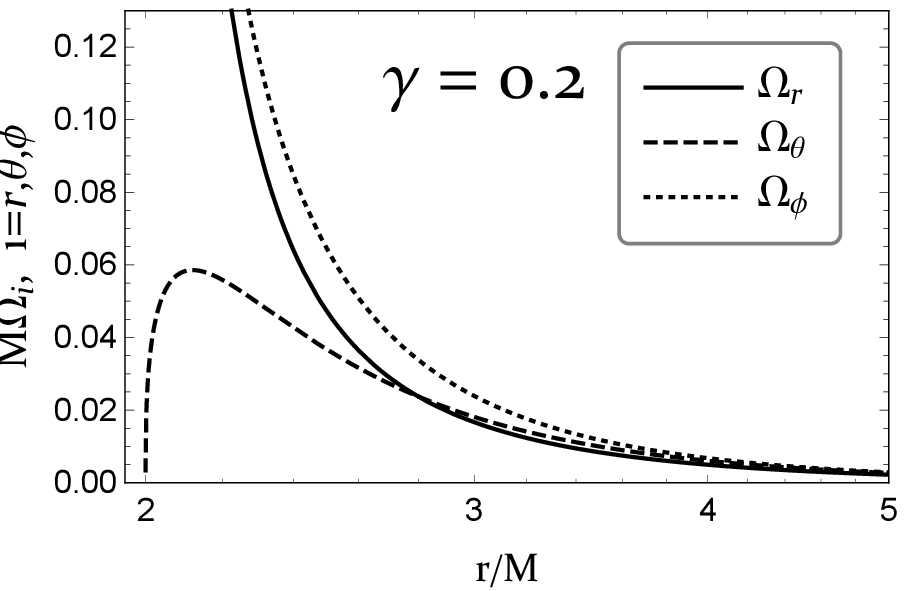}
\includegraphics[width=0.33\textwidth]{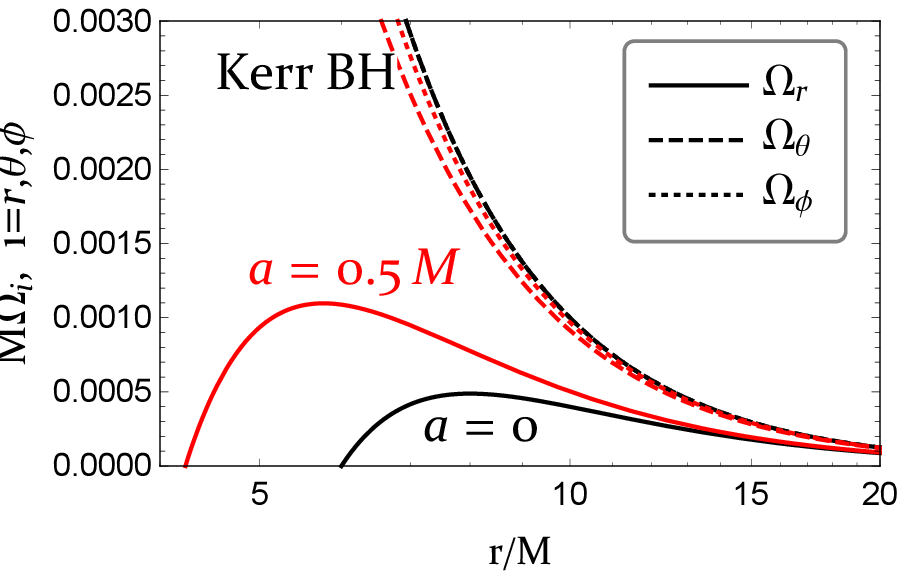}
\includegraphics[width=0.33\textwidth]{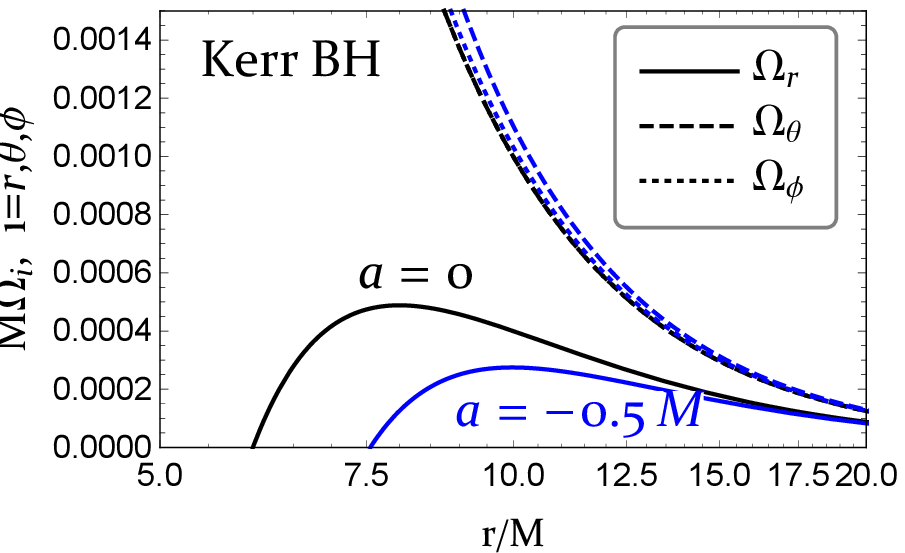}
\caption{\label{fig-freq} (color online) Radial profiles of
epicyclic frequencies for neutral test particles measured by
distant observers in the $\gamma$ space-time for different values
of the parameter $\gamma$. The last two figures represent the
radial profiles of epicyclic frequencies measured by distant
observers for neutral test particles corotating $a/M=0.5$ (red)
and counter-rotating $a/M=-0.5$ (blue) around Kerr and
Schwarzschild $a=0$ (black) black holes.}
\end{figure*}
In Fig.~\ref{fig-freq} we show the dependence of the epicyclic
frequencies for neutral test particles about the radii of stable
circular orbits for different values of $\gamma$. One can see that
the azimuthal frequencies are monotonically decreasing functions
of the radius. More in detail, depending on the value of the
parameter $\gamma$ the epicyclic frequencies as measured by
observers at spatial infinity behave as follows:

\begin{itemize}

\item[(i)] If $\gamma>1$, the latitudinal and azimuthal
epicyclic frequencies diverge, $\Omega_\theta,\Omega_\phi\rightarrow\infty$,
at $r=M(1+\gamma)$; the radial epicyclic frequency vanishes at the radius
of the outer ISCO, namely $\Omega_r=0$ at $r=r_{isco,+}$
(see first panel of Fig.~\ref{fig-freq}).

\item[(ii)] If $\gamma=1$, the latitudinal and azimuthal epicyclic
frequencies diverge, $\Omega_\theta,\Omega_\phi\rightarrow\infty$,
at $r=0$; the radial epicyclic frequency vanishes at the radius of ISCO,
namely $\Omega_r=0$ at $r=r_{isco,+}\equiv6M$. These are the known
results for the Schwarzschild space-time.

\item[(iii)] If $1/2<\gamma<1$, the latitudinal and azimuthal epicyclic
frequencies vanish, $\Omega_\theta=\Omega_\phi=0$, at $r=2M$; the radial
epicyclic frequency vanishes at the radius of ISCO, namely $\Omega_r=0$,
at $r=r_{isco,+}$ (see second panel of Fig.~\ref{fig-freq}).

\item[(iv)] If $1/\sqrt{5}\leq\gamma\leq1/2$, the latitudinal frequency
vanishes, $\Omega_\theta=0$ at $r=2M$, while the azimuthal frequency diverges,
$\Omega_\phi\rightarrow\infty$ at $r=2M$; the radial frequency vanishes at the
extremal ISCO, namely $\Omega_r=0$ at $r=(1+3/\sqrt{5})M$, and then monotonically
increases with smaller radii to diverge at the singularity, namely
$\Omega_r\rightarrow\infty$ at $r=2M$ (see third panel of Fig.~\ref{fig-freq}).

\item[(v)] If $0<\gamma<1/\sqrt{5}$, the latitudinal epicyclic frequency
vanishes, $\Omega_\theta=0$, at $r=2M$, while the radial and azimuthal
epicyclic frequencies diverge, $\Omega_r,\Omega_\phi\rightarrow\infty$,
at $r=2M$. Notice that there is no ISCO in this range of values for
$\gamma$ (see fourth panel of Fig.~\ref{fig-freq}).

\end{itemize}
Small changes in value of the parameter $\gamma$ do not affect
much the azimuthal, $\Omega_\phi$, and latitudinal,
$\Omega_\theta$, frequencies, however, the effects on the radial
frequency, $\Omega_r$, are more significant. However, a striking
difference appears in the longitudinal and azimuthal frequencies,
when we compare with the Schwarzschild case. In fact, for
$\gamma=1$ both $\Omega_\theta$ and $\Omega_\phi$ have finite, non
vanishing value at $r=2M$, while for $\gamma< 1$ they either
diverge or vanish.

The local extrema of the epicyclic frequencies are given by the
condition \bear \frac{\partial \Omega_i}{\partial r}=0 \ , \ear
where $ i=r,\theta,\phi$. For the values $\gamma\geq1/\sqrt{5}$
radial frequency always has local maximum close to the outer ISCO
radius, $r_{isco,+}$, and this maximum value increases with
decreasing value of $\gamma$. Thus, when $\gamma=1/\sqrt{5}$, the
radial epicyclic frequency of the $\gamma$ spacetime reaches the
possible highest maximum value, $\Omega_r=0.0044/M$ at
$r=(1+3/\sqrt{5})M$.

It is well known that at large distances the Schwarzschild
space-time has Newtonian behaviour and for Newton all the three
components of epicyclic frequencies have the same value given
by~\cite{Boshkayev} \bear \Omega_{r}^2=\Omega_{\theta}^2=
\Omega_{\phi}^2=\frac{M}{r^3}\ . \ear As expected, at large
distances the epicyclic frequencies in the $\gamma$ metric exhibit
Newtonian behaviour as \bear
&&\Omega_r^2=\Omega_\theta^2=\Omega_\phi^2
=\frac{M\gamma}{r^3}+O\left(\frac{1}{r^4}\right)\ , \ear which can
be interpreted as a further confirmation that the total mass of
the $\gamma$ space-time as measured by distant observers is
$M_{\rm tot}=M\gamma$.

In order to estimate the effects due to the departure from
spherical symmetry of the space-time, we can study how small
deviations from the Schwarzschild metric affect the epicyclic
frequencies. By considering $\gamma=1+\epsilon$ with
$|\epsilon|\ll1$, the epicyclic frequencies take the form
\begin{widetext}
\bear
&&\Omega_r^2=\Omega_{r,Schw}^2+\frac{M}{r^4}\left[4(r-6M)
\log\left(1-\frac{M}{r}\right)+\frac{r^2-13Mr+20 M^2}{r-2M}\right]
\epsilon+O(\epsilon^2)\ ,\label{omega-r-gamma-small}\\
&&\Omega_\theta^2=\Omega_{\theta,Schw}^2+ \frac{M}{r^3}
\left[4\log\left(1-\frac{M}{r}\right) +\frac{r-M}{r-2M}\right]
\epsilon+O(\epsilon^2)\ ,\label{omega-theta-gamma-small}\\
&&\Omega_\phi^2=\Omega_{\phi,Schw}^2+\frac{M}{r^3}
\left[2\log\left(1-\frac{2M}{r}\right) +\frac{r-M}{r-2M}\right]
\epsilon+O(\epsilon^2)\ ,\label{omega-phi-gamma-small}
\ear
\end{widetext}
where
\bear
&&\Omega_{r,Schw}^2=\frac{M(r-6M)}{r^4}\ ,\\
&&\Omega_{\theta,Schw}^2=\Omega_{\phi,Schw}^2=\frac{M}{r^3}\ ,
\ear Thus, one can see from
Eqs.~(\ref{omega-r-gamma-small}),~(\ref{omega-theta-gamma-small}),
 (\ref{omega-phi-gamma-small}) that oblate (prolate), i.e. values of $\epsilon>0$ ($\epsilon<0$),
deviations from spherical symmetry, i.e. Schwarzschild, increase
the values of all components of the epicyclic frequency of test
particles.

\subsection{Epicyclic frequencies in Kerr spacetime}

In order to compare epicyclic frequencies of neutral test
particles  moving along circular orbits in the $\gamma$ space-time
with the ones in the Kerr space-time, we briefly review the
epicyclic frequencies around a Kerr black hole. In Boyer-Lindquist
coordinates the line element for the Kerr space-time is given by:
\bear\label{kerr-spacetime}
ds^2&&=-\left(1-\frac{2Mr}{\Sigma}\right)dt^2+\frac{\Sigma}{\Delta}dr^2
-2\frac{2Mra}{\Sigma}\sin^2\theta d\phi dt\nonumber\\
&&+\Sigma
d\theta^2+\left(r^2+a^2+\frac{2Mra^2}{\Sigma}\sin^2\theta\right)d\phi^2\
, \ear where \bear \Sigma=r^2+a^2\cos^2\theta,\qquad
\Delta=r^2-2Mr+a^2 \ . \ear The comparison between the $\gamma$
and the Kerr space-times is justified by the fact at large
distances both metrics tend to become asymptotically flat and so
Erez-Rosen as well as Boyer-Lindquist coordinates tend to become
the usual spherical coordinates. Therefore an observer at
infinity, measuring some features of the motion of test particles
in the accretion disk around a Kerr black hole would be able to
compare, at least in principle, the obtained measurements with the
expected observations of the corresponding situation in the
$\gamma$ space-time. Similarly to the static axially symmetric
case, in the stationary case, due to rotation of the source,
circular orbits are located on the equatorial plane, thus allowing
us to restrict the analysis to the case $\theta_0=\pi/2$. The
specific energy, angular momentum and angular velocity of the
particle moving on a circular orbit around a Kerr black hole are
derived from relations~(\ref{ener}), (\ref{angul})
and~(\ref{ang-vel}), respectively as \bear
&&E_\pm=\frac{\sqrt{r}(r-2M)\pm a\sqrt{M}}{r\sqrt{r\pm2a\sqrt{M/r}-3M}}\ ,\label{kerr-ener}\\
&&L_\pm=\pm\frac{\sqrt{M}(r^2\mp2a\sqrt{M/r}+a^2)}{ r\sqrt{r\pm2a\sqrt{M/r}-3M}}\ ,\\
&&\Omega_\pm=\pm\frac{\sqrt{M}}{a\sqrt{M}\pm r\sqrt{r}}\ . \ear
Where the $+$ and $-$ signs correspond to co-rotating and
counter-rotating particles with respect to the direction of
rotation of the black hole. Again one can find the radii of
circular photon orbits, marginally bound orbits and ISCO from the
following relations~\cite{Bardeen:APJ:1972}: \bear
&&r_{ps\pm}\pm2a\sqrt{M/r_{ps\pm}}-3M=0\ ,\\
&&r_{mb\pm}=2M\pm a+2\sqrt{M(M\pm a)}\ ,\\
&&r_{isco\pm}(r_{isco\pm}-6M)\pm8ar_{isco\pm} \sqrt{M/r_{isco\pm}}-3a^2=0,\nonumber\\
\ear
\begin{figure*}[th]
\centering
\includegraphics[width=0.48\textwidth]{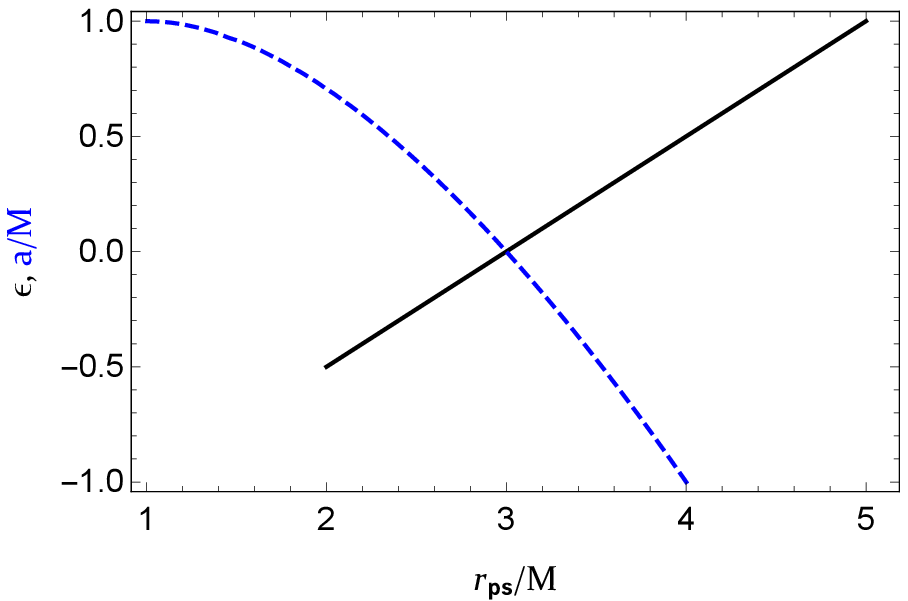}
\includegraphics[width=0.48\textwidth]{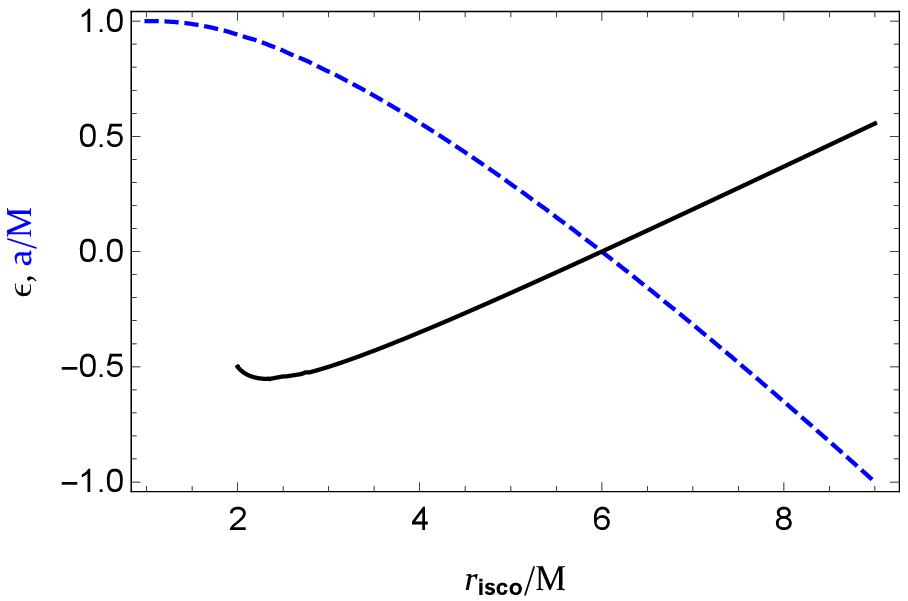}
\caption{\label{fig-kerr-orbits} (Color online) Left panel: Radii
of circular photon orbits in $\gamma$ spacetime (black) with
$\gamma=1+\epsilon$ and Kerr black hole (blue, dashed). Right
panel: ISCO radii of neutral test particles in $\gamma$ spacetime
(black) and Kerr black hole (blue, dashed).}
\end{figure*}

In Fig.~\ref{fig-kerr-orbits} we show the degeneracy between the
$\gamma$ metric and the Kerr space-time resulting from
measurements of the ISCO or of the photon capture radius. In fact,
each given measured value of $r_{isco}$ corresponds to one value
of $a$ and one value of $\gamma$ showing that the measurement of
the ISCO radius alone is not enough to distinguish between the two
space-times, unless the value is smaller than $2M$, as the
$\gamma$ space-time cannot have characteristic orbits whose radii
are smaller than $2M$, while rapidly rotating Kerr can. Similar
arguments hold for the photon capture radius. However, by
observing that the degeneracy in the two cases is different we can
conclude that a simultaneous measurement of $r_{isco}$ and
$r_{ps}$ with enough precision would allow to distinguish between
the two geometries around astrophysical compact objects.

The epicyclic frequencies, with respect to distant observers, of a
test particle moving along circular orbits around a Kerr black
hole are given by~\cite{Torok:AA:2005,Torok:AA:2005a} \bear
&&\Omega_{r\pm}^2=\frac{M\left[r(r-6M)\pm8ar \sqrt{M/r}-3a^2\right]}{r^2\left(r^3\pm2ar^2\sqrt{M/r}+a^2M\right)}\ ,\\
&&\Omega_{\theta\pm}^2=\frac{M\left(r^2\mp4ar \sqrt{M/r}+3a^2\right)}{r^2\left(r^3\pm2ar^2\sqrt{M/r}+a^2M\right)}\ ,\\
&&\Omega_{\phi\pm}^2=\frac{M}{r^3\pm2ar^2\sqrt{M/r}+a^2M}\ .
\ear
In the slow rotation limit, $a\ll M$, the above expressions take the form
\bear
&&\Omega_{r\pm}^2=\Omega_{r,Schw}^2\pm \frac{6\left(M/r\right)^{3/2}(r+2M)}{r^4}a+O(a^2),\label{omega-r-kerr-slow}\\
&&\Omega_{\theta\pm}^2=\Omega_{\theta,Schw}^2\mp \frac{6\left(M/r\right)^{3/2}}{r^3}a+O(a^2)\ ,\label{omega-theta-kerr-slow}\\
&&\Omega_{\phi\pm}^2=\Omega_{\phi,Schw}^2\mp
\frac{2\left(M/r\right)^{3/2}}{r^3}a+O(a^2)\
.\label{omega-phi-kerr-slow} \ear Thus, one can see
from~(\ref{omega-r-kerr-slow}),~(\ref{omega-theta-kerr-slow}),~(\ref{omega-phi-kerr-slow})
that rotation of the spacetime increases (decreases) radial
frequency of the harmonic oscillations and decreases (increases)
latitudinal and azimuthal frequency of the harmonic oscillations
of the co-rotating (counter-rotating) particle moving along the
circular orbit.

This shows that the degeneracy between the Kerr metric and the
$\gamma$ metric due to measurements of the value of the ISCO can
be broken if one is able to measure simultaneously $\Omega_r$ and
one between $\Omega_\theta$ and $\Omega_\phi$.

\begin{figure}[h]
\centering
\includegraphics[width=0.48\textwidth]{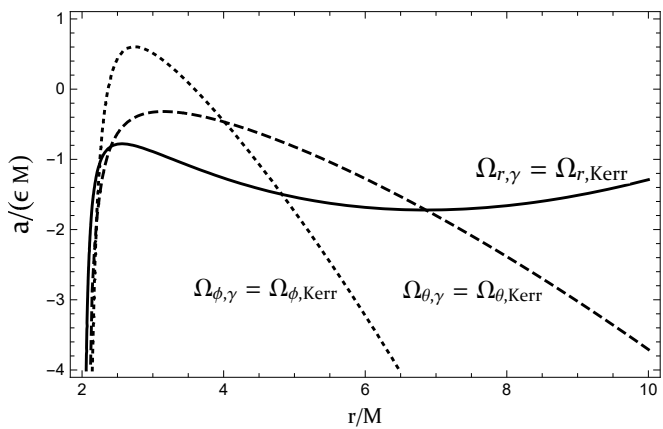}
\caption{\label{fig-kerr-gamma} Radial dependence of ratio
$a/\epsilon$ for which the $\gamma$ and Kerr space-times have the
same epicyclic frequencies. }
\end{figure}

To this aim, we consider the values of $\gamma$ and $a$ for which
the epicyclic frequencies of the neutral test particles have the
same value. As we mentioned, if the circular orbits where the
particles are moving are located far away from the central object
then the two metrics can not be distinguished, since at large
distances all frequencies become Newtonian. Therefore, we will
focus our attention on relatively short distances. For simplicity
we shall also consider slight deviations from the Schwarzschild
space-time for both cases, i.e. small deformations
$\gamma=1+\epsilon$ with $|\epsilon|\ll1$, and slow rotation,
$|a|\ll M$. Thus, by using the expressions of epicyclic
frequencies for the $\gamma$-metric given by
Eqs.~(\ref{omega-r-gamma-small}),~(\ref{omega-theta-gamma-small}),~(\ref{omega-phi-gamma-small})
and for the Kerr black hole, given by
Eqs.~(\ref{omega-r-kerr-slow}),~(\ref{omega-theta-kerr-slow}),~(\ref{omega-phi-kerr-slow})
we find the expressions for the coincidence of all components of
the frequencies as
\begin{widetext}
\bear
&&\Omega_{r,\gamma}^2=\Omega_{r,Kerr}^2\quad\Rightarrow\quad \frac{a}{\epsilon}=
\frac{M\left[4 (r-6M)(r-2M)\log\left(1-M/r\right)+r^2-13Mr+20 M^2\right]}{
\left(r^2-4M^2\right)}\left(\frac{r}{M}\right)^{3/2}\ ,\\
&&\Omega_{\theta,\gamma}^2=\Omega_{\theta,Kerr}^2\quad\Rightarrow\quad
\frac{a}{\epsilon}=-\frac{M}{6} \left[4\log\left(1-\frac{M}{r}\right)+
\frac{r-M}{r-2 M}\right]\left(\frac{r}{M}\right)^{3/2}\ ,\\
&&\Omega_{\phi,\gamma}^2=\Omega_{\phi,Kerr}^2\quad\Rightarrow\quad
\frac{a}{\epsilon}=-\frac{M}{2} \left[2 \log \left(1-\frac{2 M}{r}\right)
+\frac{r-M}{r-2 M}\right]\left(\frac{r}{M}\right)^{3/2}\ .
\ear
\end{widetext}
In Fig.~\ref{fig-kerr-gamma} we show the above expressions as
functions of the dimensionless radial coordinate $r/M$.

Since test particles are oscillating due to slight deviations from
stable circular orbits in a space-time which slightly deviates
from spherically symmetry, it is useful to focus the attention on
the region near $r/M\approx6$, corresponding to the ISCO for the
Schwarzschild metric. One can see from Fig.~\ref{fig-kerr-gamma}
that the three curves corresponding to the same values for the
three frequencies in Kerr and in the $\gamma$ metric intersect in
three different points. Therefore, if one measurement was in
principle able to measure all three frequencies, the results could
easily distinguish the $\gamma$ space-time from Kerr. On the other
hand, a measurement of only two frequencies, may not be enough if
performed in the vicinity of the point where two curves overlap.
For example, the curves obtained from $\Omega_r$ and
$\Omega_\theta$ overlap at $r\approx6.8M$ with
$a/\epsilon\approx-1.7M$. Thus, a measurement of these two
frequencies in the vicinity of this radius would not be able to
distinguish the $\gamma$ space-time from Kerr.

From the above considerations we can derive some conclusion on
which kind of objects may or may not be distinguished from a Kerr
black hole.
\begin{itemize}

\item[(i)] If test particles are co-rotating relative to the Kerr
black hole ($aL>0$) and the source of the $\gamma$ space-time is
oblate ($\epsilon>0$), then the two geometries are distinguishable.

\item[(ii)] If test particles are co-rotating relative to the Kerr
black hole and the source of the $\gamma$ space-time is prolate
($\epsilon<0$), then the two geometries may be indistinguishable.

\item[(iii)] If test particles are counter-rotating relative to
the Kerr black hole ($aL<0$) and the source of the $\gamma$
space-time is oblate, then the two geometries may be indistinguishable.

\item[(iv)] If test particles are counter-rotating relative to
the Kerr black hole and the source of the $\gamma$ space-time
is prolate, then the two geometries are distinguishable.

\end{itemize}
Finally, if the latitudinal and radial frequencies of harmonic
oscillations of a particle moving along the circular orbit with
radius $r=6.8M$ are detected, one will not be able to distinguish
if the central gravitating object is described by the slowly
rotating Kerr (and which rotational direction relative to central
object is particle moving) or slightly deformed $\gamma$ (and type
of deformation) space-times. Since for all three frequencies the
ratio $a/\epsilon$ is negative, similar considerations as the ones
above apply also to the simultaneous measurement of radial and
azimuthal frequencies and to the simultaneous measurement of
azimuthal and latitudinal frequencies.

\section{Conclusion}\label{conc}

The recent observation of the ``black hole shadow" by the Event
Horizon Telescope, has opened the door to precise measurements of
the geometry in the vicinity of black hole candidates~\cite{EHT}.
To this day, astrophysical observations of black hole candidates
have shown no departure from the relativistic description (see for
example the recent measurement of the gravitational redshift of
stars orbiting the supermassive black hole candidate at the center
of the Milky Way~\cite{AAP}). However such measurements have not
been able yet to test whether the geometry in the vicinity of
black hole candidates is well described by the Kerr metric. For
this reason it is useful to study the predictions obtained from
different geometries which, while being solutions of Einstein's
field equations in vacuum, do not describe a black hole. In this
respect the $\gamma$ metric is an ideal candidate due to its
simplicity and to its immediate connection to the Schwarzschild
space-time.

The $\gamma$ metric describes a deformed Schwarzschild object,  it
is prolate for $\gamma>1$ and oblate for $\gamma<1$. It turns out
that for the former, particle dynamics continues to have the same
qualitative features as the Schwarzschild's while for the latter
it has much richer structure. In particular, there are two
disjoint regions for existence of stable circular orbits for the
range $1/\sqrt{5} < \gamma < 1/2$ while for $\gamma < 1/\sqrt{5}$,
there exist only stable circular orbits. This is exactly like the
Newtonian case.

In the present paper we have studied some properties of
characteristic circular orbits, i.e. circular null geodesics,
marginally bound orbits, stable circular orbits, for massive and
massless test particles around the $\gamma$ metric. In particular
we focused on the frequencies of harmonic oscillations around
stable circular orbits for massive test particles.

We have shown that small deviations from equatorial circular
orbits of the particle in the $\gamma$ space-time are described by
harmonic oscillations in the radial, vertical, and latitudinal
directions. By solving the harmonic oscillator equations we have
found the analytical expressions for the frequencies and studied
the epicyclic frequencies of particles in the uncoupled orthogonal
(radial), vertical (latitudinal) and axial (azimuthal) oscillatory
modes relative to comoving and distant observers. The frequencies
as measured by distant observers are particularly important since
they could, in principle, be measured from the observations of
accretion disks around black hole candidates.

We have shown that radial oscillations vanish at the ISCO of the
space-time, and below the ISCO the particles fall towards the
central object, and no radial oscillations occur. Since at large
distances the space-time approaches the Newtonian limit, all
epicyclic frequencies tend to the same limit
$\Omega_i^2=M\gamma/r^3$ with $i=r,\theta,\phi$. For this reason,
we have studied the behaviour of the frequencies in the
relativistic regime for slight deviations from spherical symmetry
and compared the results with the corresponding frequencies in the
slowly rotating Kerr black hole geometry.

We have shown that the two metrics are distinguishable if a
simultaneous measurement of all three components of the epicyclic
frequencies is available. However, if only two frequencies of
oscillations for the test particles are obtained (for example near
$r\approx6.8M$ for the radial and vertical oscillations) then it
may not be possible to distinguish if the geometry around the
central object is described by the $\gamma$ or Kerr space-times.
Similarly, we have shown that a simultaneous measurement of the
radii of the ISCO and the photon capture orbit may be able to
distinguish between the two geometries, while the measurement of
one radius only is not enough unless such radius is smaller than
$2M$. This is because rapidly rotating Kerr black hole can have
orbits below $2M$ while the $\gamma$ spacetime cannot.

As new an better observations of astrophysical black hole
candidates become available, we will soon be able to test the
nature of the geometry around such candidates and answer the
question whether all such objects must necessarily be described by
the Kerr space-time or if nature allows for some other
possibility.

\section*{Acknowledgments}

The work was developed under the Nazarbayev University Faculty
Development Competitive Research Grant No.~090118FD5348. The
authors acknowledge the support of the Ministry of Education of
Kazakhstan's target program IRN:~BR05236454 and Uzbekistan
Ministry for Innovation Development Grants No.~VA-FA-F-2-008 and
No.~YFA-Ftech-2018-8. N.D. would like to acknowledge Nazarbayev
University, Astana, Kazakhstan for the hospitality and Albert
Einstein Institute, Golm, Germany for a visit that partially
facilitated this work. D.M. and N.D. wish to express their
gratitude to Bobomurat Ahmedov for useful discussion on the
properties of the gamma metric and acknowledge the Ulugh Beg
Astronomical Institute, Tashkent, Uzbekistan for the hospitality
through Abdus Salam International Centre for Theoretical Physics
through Grant No.~OEA-NT-01.

\label{lastpage}

\bibliography{gamma_references}

\begin{thebibliography}{35}%
\makeatletter
\providecommand \@ifxundefined [1]{%
 \@ifx{#1\undefined}
}%
\providecommand \@ifnum [1]{%
 \ifnum #1\expandafter \@firstoftwo
 \else \expandafter \@secondoftwo
 \fi
}%
\providecommand \@ifx [1]{%
 \ifx #1\expandafter \@firstoftwo
 \else \expandafter \@secondoftwo
 \fi
}%
\providecommand \natexlab [1]{#1}%
\providecommand \enquote  [1]{``#1''}%
\providecommand \bibnamefont  [1]{#1}%
\providecommand \bibfnamefont [1]{#1}%
\providecommand \citenamefont [1]{#1}%
\providecommand \href@noop [0]{\@secondoftwo}%
\providecommand \href [0]{\begingroup \@sanitize@url \@href}%
\providecommand \@href[1]{\@@startlink{#1}\@@href}%
\providecommand \@@href[1]{\endgroup#1\@@endlink}%
\providecommand \@sanitize@url [0]{\catcode `\\12\catcode `\$12\catcode
  `\&12\catcode `\#12\catcode `\^12\catcode `\_12\catcode `\%12\relax}%
\providecommand \@@startlink[1]{}%
\providecommand \@@endlink[0]{}%
\providecommand \url  [0]{\begingroup\@sanitize@url \@url }%
\providecommand \@url [1]{\endgroup\@href {#1}{\urlprefix }}%
\providecommand \urlprefix  [0]{URL }%
\providecommand \Eprint [0]{\href }%
\providecommand \doibase [0]{http://dx.doi.org/}%
\providecommand \selectlanguage [0]{\@gobble}%
\providecommand \bibinfo  [0]{\@secondoftwo}%
\providecommand \bibfield  [0]{\@secondoftwo}%
\providecommand \translation [1]{[#1]}%
\providecommand \BibitemOpen [0]{}%
\providecommand \bibitemStop [0]{}%
\providecommand \bibitemNoStop [0]{.\EOS\space}%
\providecommand \EOS [0]{\spacefactor3000\relax}%
\providecommand \BibitemShut  [1]{\csname bibitem#1\endcsname}%
\let\auto@bib@innerbib\@empty
\bibitem [{\citenamefont {{Zipoy}}(1966)}]{Zip}%
  \BibitemOpen
  \bibfield  {author} {\bibinfo {author} {\bibfnamefont {D.~M.}\ \bibnamefont
  {{Zipoy}}},\ }\href {\doibase 10.1063/1.1705005} {\bibfield  {journal}
  {\bibinfo  {journal} {J. Math. Phys.}\ }\textbf {\bibinfo {volume} {7}},\
  \bibinfo {pages} {1137} (\bibinfo {year} {1966})}\BibitemShut {NoStop}%
\bibitem [{\citenamefont {{Voorhees}}(1970)}]{Vor}%
  \BibitemOpen
  \bibfield  {author} {\bibinfo {author} {\bibfnamefont {B.~H.}\ \bibnamefont
  {{Voorhees}}},\ }\href {\doibase 10.1103/PhysRevD.2.2119} {\bibfield
  {journal} {\bibinfo  {journal} {Phys. Rev. D}\ }\textbf {\bibinfo {volume}
  {2}},\ \bibinfo {pages} {2119} (\bibinfo {year} {1970})}\BibitemShut
  {NoStop}%
\bibitem [{\citenamefont {{Virbhadra}}(1996)}]{virb}%
  \BibitemOpen
  \bibfield  {author} {\bibinfo {author} {\bibfnamefont {K.~S.}\ \bibnamefont
  {{Virbhadra}}},\ }\href@noop {} {\bibfield  {journal} {\bibinfo  {journal}
  {arXiv e-prints}\ ,\ \bibinfo {eid} {gr-qc/9606004}} (\bibinfo {year}
  {1996})},\ \Eprint {http://arxiv.org/abs/gr-qc/9606004} {arXiv:gr-qc/9606004
  [astro-ph]} \BibitemShut {NoStop}%
\bibitem [{\citenamefont {{Papadopoulos}}\ \emph {et~al.}(1981)\citenamefont
  {{Papadopoulos}}, \citenamefont {{Stewart}},\ and\ \citenamefont
  {{Witten}}}]{Papadopoulos:PRD:1981}%
  \BibitemOpen
  \bibfield  {author} {\bibinfo {author} {\bibfnamefont {D.}~\bibnamefont
  {{Papadopoulos}}}, \bibinfo {author} {\bibfnamefont {B.}~\bibnamefont
  {{Stewart}}}, \ and\ \bibinfo {author} {\bibfnamefont {L.}~\bibnamefont
  {{Witten}}},\ }\href {\doibase 10.1103/PhysRevD.24.320} {\bibfield  {journal}
  {\bibinfo  {journal} {Phys. Rev. D}\ }\textbf {\bibinfo {volume} {24}},\
  \bibinfo {pages} {320} (\bibinfo {year} {1981})}\BibitemShut {NoStop}%
\bibitem [{\citenamefont {{Herrera}}\ \emph {et~al.}(1999)\citenamefont
  {{Herrera}}, \citenamefont {{Paiva}},\ and\ \citenamefont
  {{Santos}}}]{herrera1}%
  \BibitemOpen
  \bibfield  {author} {\bibinfo {author} {\bibfnamefont {L.}~\bibnamefont
  {{Herrera}}}, \bibinfo {author} {\bibfnamefont {F.~M.}\ \bibnamefont
  {{Paiva}}}, \ and\ \bibinfo {author} {\bibfnamefont {N.~O.}\ \bibnamefont
  {{Santos}}},\ }\href {\doibase 10.1063/1.532943} {\bibfield  {journal}
  {\bibinfo  {journal} {J. Math. Phys.}\ }\textbf {\bibinfo {volume} {40}},\
  \bibinfo {pages} {4064} (\bibinfo {year} {1999})},\ \Eprint
  {http://arxiv.org/abs/gr-qc/9810079} {gr-qc/9810079} \BibitemShut {NoStop}%
\bibitem [{\citenamefont {{Herrera}}\ and\ \citenamefont
  {{Pastora}}(2000)}]{herrera2}%
  \BibitemOpen
  \bibfield  {author} {\bibinfo {author} {\bibfnamefont {L.}~\bibnamefont
  {{Herrera}}}\ and\ \bibinfo {author} {\bibfnamefont {J.~L.~H.}\ \bibnamefont
  {{Pastora}}},\ }\href {\doibase 10.1063/1.1319517} {\bibfield  {journal}
  {\bibinfo  {journal} {J. Math. Phys.}\ }\textbf {\bibinfo {volume} {41}},\
  \bibinfo {pages} {7544} (\bibinfo {year} {2000})},\ \Eprint
  {http://arxiv.org/abs/gr-qc/0010003} {gr-qc/0010003} \BibitemShut {NoStop}%
\bibitem [{\citenamefont {{Hern{\'a}ndez-Pastora}}\ and\ \citenamefont
  {{Mart{\'{\i}}n}}(1994)}]{hernandez}%
  \BibitemOpen
  \bibfield  {author} {\bibinfo {author} {\bibfnamefont {J.~L.}\ \bibnamefont
  {{Hern{\'a}ndez-Pastora}}}\ and\ \bibinfo {author} {\bibfnamefont
  {J.}~\bibnamefont {{Mart{\'{\i}}n}}},\ }\href {\doibase 10.1007/BF02107146}
  {\bibfield  {journal} {\bibinfo  {journal} {Gen. Rel. Grav.}\ }\textbf
  {\bibinfo {volume} {26}},\ \bibinfo {pages} {877} (\bibinfo {year}
  {1994})}\BibitemShut {NoStop}%
\bibitem [{\citenamefont {{Bonnor}}(1992)}]{Bonnor:GRG:1992}%
  \BibitemOpen
  \bibfield  {author} {\bibinfo {author} {\bibfnamefont {W.~B.}\ \bibnamefont
  {{Bonnor}}},\ }\href {\doibase 10.1007/BF00760137} {\bibfield  {journal}
  {\bibinfo  {journal} {General Relativity and Gravitation}\ }\textbf {\bibinfo
  {volume} {24}},\ \bibinfo {pages} {551} (\bibinfo {year} {1992})}\BibitemShut
  {NoStop}%
\bibitem [{\citenamefont {{Hernandez}}(1967)}]{interior0}%
  \BibitemOpen
  \bibfield  {author} {\bibinfo {author} {\bibfnamefont {W.~C.}\ \bibnamefont
  {{Hernandez}}},\ }\href {\doibase 10.1103/PhysRev.153.1359} {\bibfield
  {journal} {\bibinfo  {journal} {Phys. Rev.}\ }\textbf {\bibinfo {volume}
  {153}},\ \bibinfo {pages} {1359} (\bibinfo {year} {1967})}\BibitemShut
  {NoStop}%
\bibitem [{\citenamefont {{Stewart}}\ \emph {et~al.}(1982)\citenamefont
  {{Stewart}}, \citenamefont {{Papadopoulos}}, \citenamefont {{Witten}},
  \citenamefont {{Berezdivin}},\ and\ \citenamefont {{Herrera}}}]{interior1}%
  \BibitemOpen
  \bibfield  {author} {\bibinfo {author} {\bibfnamefont {B.~W.}\ \bibnamefont
  {{Stewart}}}, \bibinfo {author} {\bibfnamefont {D.}~\bibnamefont
  {{Papadopoulos}}}, \bibinfo {author} {\bibfnamefont {L.}~\bibnamefont
  {{Witten}}}, \bibinfo {author} {\bibfnamefont {R.}~\bibnamefont
  {{Berezdivin}}}, \ and\ \bibinfo {author} {\bibfnamefont {L.}~\bibnamefont
  {{Herrera}}},\ }\href {\doibase 10.1007/BF00756201} {\bibfield  {journal}
  {\bibinfo  {journal} {Gen. Rel. Grav.}\ }\textbf {\bibinfo {volume} {14}},\
  \bibinfo {pages} {97} (\bibinfo {year} {1982})}\BibitemShut {NoStop}%
\bibitem [{\citenamefont {{Herrera}}\ \emph {et~al.}(2005)\citenamefont
  {{Herrera}}, \citenamefont {{Magli}},\ and\ \citenamefont
  {{Malafarina}}}]{interior2}%
  \BibitemOpen
  \bibfield  {author} {\bibinfo {author} {\bibfnamefont {L.}~\bibnamefont
  {{Herrera}}}, \bibinfo {author} {\bibfnamefont {G.}~\bibnamefont {{Magli}}},
  \ and\ \bibinfo {author} {\bibfnamefont {D.}~\bibnamefont {{Malafarina}}},\
  }\href {\doibase 10.1007/s10714-005-0120-1} {\bibfield  {journal} {\bibinfo
  {journal} {Gen. Rel. Grav.}\ }\textbf {\bibinfo {volume} {37}},\ \bibinfo
  {pages} {1371} (\bibinfo {year} {2005})},\ \Eprint
  {http://arxiv.org/abs/gr-qc/0407037} {gr-qc/0407037} \BibitemShut {NoStop}%
\bibitem [{\citenamefont {{Chowdhury}}\ \emph {et~al.}(2012)\citenamefont
  {{Chowdhury}}, \citenamefont {{Patil}}, \citenamefont {{Malafarina}},\ and\
  \citenamefont {{Joshi}}}]{disk}%
  \BibitemOpen
  \bibfield  {author} {\bibinfo {author} {\bibfnamefont {A.~N.}\ \bibnamefont
  {{Chowdhury}}}, \bibinfo {author} {\bibfnamefont {M.}~\bibnamefont
  {{Patil}}}, \bibinfo {author} {\bibfnamefont {D.}~\bibnamefont
  {{Malafarina}}}, \ and\ \bibinfo {author} {\bibfnamefont {P.~S.}\
  \bibnamefont {{Joshi}}},\ }\href {\doibase 10.1103/PhysRevD.85.104031}
  {\bibfield  {journal} {\bibinfo  {journal} {Phys. Rev. D}\ }\textbf {\bibinfo
  {volume} {85}},\ \bibinfo {eid} {104031} (\bibinfo {year} {2012})},\ \Eprint
  {http://arxiv.org/abs/1112.2522} {arXiv:1112.2522 [gr-qc]} \BibitemShut
  {NoStop}%
\bibitem [{\citenamefont {{Boshkayev}}\ \emph {et~al.}(2016)\citenamefont
  {{Boshkayev}}, \citenamefont {{Gasper{\'\i}n}}, \citenamefont
  {{Guti{\'e}rrez-Pi{\~n}eres}}, \citenamefont {{Quevedo}},\ and\ \citenamefont
  {{Toktarbay}}}]{Boshkayev:PRD:2016}%
  \BibitemOpen
  \bibfield  {author} {\bibinfo {author} {\bibfnamefont {K.}~\bibnamefont
  {{Boshkayev}}}, \bibinfo {author} {\bibfnamefont {E.}~\bibnamefont
  {{Gasper{\'\i}n}}}, \bibinfo {author} {\bibfnamefont {A.~C.}\ \bibnamefont
  {{Guti{\'e}rrez-Pi{\~n}eres}}}, \bibinfo {author} {\bibfnamefont
  {H.}~\bibnamefont {{Quevedo}}}, \ and\ \bibinfo {author} {\bibfnamefont
  {S.}~\bibnamefont {{Toktarbay}}},\ }\href {\doibase
  10.1103/PhysRevD.93.024024} {\bibfield  {journal} {\bibinfo  {journal} {Phys.
  Rev. D}\ }\textbf {\bibinfo {volume} {93}},\ \bibinfo {eid} {024024}
  (\bibinfo {year} {2016})},\ \Eprint {http://arxiv.org/abs/1509.03827}
  {arXiv:1509.03827 [gr-qc]} \BibitemShut {NoStop}%
\bibitem [{\citenamefont {{Benavides-Gallego}}\ \emph
  {et~al.}(2019)\citenamefont {{Benavides-Gallego}}, \citenamefont
  {{Abdujabbarov}}, \citenamefont {{Malafarina}}, \citenamefont {{Ahmedov}},\
  and\ \citenamefont {{Bambi}}}]{us1}%
  \BibitemOpen
  \bibfield  {author} {\bibinfo {author} {\bibfnamefont {C.~A.}\ \bibnamefont
  {{Benavides-Gallego}}}, \bibinfo {author} {\bibfnamefont {A.}~\bibnamefont
  {{Abdujabbarov}}}, \bibinfo {author} {\bibfnamefont {D.}~\bibnamefont
  {{Malafarina}}}, \bibinfo {author} {\bibfnamefont {B.}~\bibnamefont
  {{Ahmedov}}}, \ and\ \bibinfo {author} {\bibfnamefont {C.}~\bibnamefont
  {{Bambi}}},\ }\href {\doibase 10.1103/PhysRevD.99.044012} {\bibfield
  {journal} {\bibinfo  {journal} {Phys. Rev. D}\ }\textbf {\bibinfo {volume}
  {99}},\ \bibinfo {eid} {044012} (\bibinfo {year} {2019})},\ \Eprint
  {http://arxiv.org/abs/1812.04846} {arXiv:1812.04846 [gr-qc]} \BibitemShut
  {NoStop}%
\bibitem [{\citenamefont {{Abdikamalov}}\ \emph {et~al.}(2019)\citenamefont
  {{Abdikamalov}}, \citenamefont {{Abdujabbarov}}, \citenamefont {{Ayzenberg}},
  \citenamefont {{Malafarina}}, \citenamefont {{Bambi}},\ and\ \citenamefont
  {{Ahmedov}}}]{Askar:arxiv:2019}%
  \BibitemOpen
  \bibfield  {author} {\bibinfo {author} {\bibfnamefont {A.~B.}\ \bibnamefont
  {{Abdikamalov}}}, \bibinfo {author} {\bibfnamefont {A.~A.}\ \bibnamefont
  {{Abdujabbarov}}}, \bibinfo {author} {\bibfnamefont {D.}~\bibnamefont
  {{Ayzenberg}}}, \bibinfo {author} {\bibfnamefont {D.}~\bibnamefont
  {{Malafarina}}}, \bibinfo {author} {\bibfnamefont {C.}~\bibnamefont
  {{Bambi}}}, \ and\ \bibinfo {author} {\bibfnamefont {B.}~\bibnamefont
  {{Ahmedov}}},\ }\href@noop {} {\bibfield  {journal} {\bibinfo  {journal}
  {arXiv e-prints}\ ,\ \bibinfo {eid} {arXiv:1904.06207}} (\bibinfo {year}
  {2019})},\ \Eprint {http://arxiv.org/abs/1904.06207} {arXiv:1904.06207
  [gr-qc]} \BibitemShut {NoStop}%
\bibitem [{\citenamefont {{Abramowicz}}\ and\ \citenamefont
  {{Klu{\'z}niak}}(2005)}]{Abramowicz:APSS:2005}%
  \BibitemOpen
  \bibfield  {author} {\bibinfo {author} {\bibfnamefont {M.~A.}\ \bibnamefont
  {{Abramowicz}}}\ and\ \bibinfo {author} {\bibfnamefont {W.}~\bibnamefont
  {{Klu{\'z}niak}}},\ }\href {\doibase 10.1007/s10509-005-1173-z} {\bibfield
  {journal} {\bibinfo  {journal} {Astrophysics and Space Science}\ }\textbf
  {\bibinfo {volume} {300}},\ \bibinfo {pages} {127} (\bibinfo {year}
  {2005})},\ \Eprint {http://arxiv.org/abs/astro-ph/0411709}
  {arXiv:astro-ph/0411709 [astro-ph]} \BibitemShut {NoStop}%
\bibitem [{\citenamefont {{Kolo{\v s}}}\ \emph {et~al.}(2015)\citenamefont
  {{Kolo{\v s}}}, \citenamefont {{Stuchl{\'{\i}}k}},\ and\ \citenamefont
  {{Tursunov}}}]{Kolos:CQG:2015}%
  \BibitemOpen
  \bibfield  {author} {\bibinfo {author} {\bibfnamefont {M.}~\bibnamefont
  {{Kolo{\v s}}}}, \bibinfo {author} {\bibfnamefont {Z.}~\bibnamefont
  {{Stuchl{\'{\i}}k}}}, \ and\ \bibinfo {author} {\bibfnamefont
  {A.}~\bibnamefont {{Tursunov}}},\ }\href {\doibase
  10.1088/0264-9381/32/16/165009} {\bibfield  {journal} {\bibinfo  {journal}
  {Classical and Quantum Gravity}\ }\textbf {\bibinfo {volume} {32}},\ \bibinfo
  {eid} {165009} (\bibinfo {year} {2015})},\ \Eprint
  {http://arxiv.org/abs/1506.06799} {arXiv:1506.06799 [gr-qc]} \BibitemShut
  {NoStop}%
\bibitem [{\citenamefont {{Kolo{\v s}}}\ \emph {et~al.}(2017)\citenamefont
  {{Kolo{\v s}}}, \citenamefont {{Tursunov}},\ and\ \citenamefont
  {{Stuchl{\'{\i}}k}}}]{Kolos:EPJC:2017}%
  \BibitemOpen
  \bibfield  {author} {\bibinfo {author} {\bibfnamefont {M.}~\bibnamefont
  {{Kolo{\v s}}}}, \bibinfo {author} {\bibfnamefont {A.}~\bibnamefont
  {{Tursunov}}}, \ and\ \bibinfo {author} {\bibfnamefont {Z.}~\bibnamefont
  {{Stuchl{\'{\i}}k}}},\ }\href {\doibase 10.1140/epjc/s10052-017-5431-3}
  {\bibfield  {journal} {\bibinfo  {journal} {Eur. Phys. J. C}\ }\textbf
  {\bibinfo {volume} {77}},\ \bibinfo {eid} {860} (\bibinfo {year} {2017})},\
  \Eprint {http://arxiv.org/abs/1707.02224} {arXiv:1707.02224 [astro-ph.HE]}
  \BibitemShut {NoStop}%
\bibitem [{\citenamefont {{Tursunov}}\ \emph {et~al.}(2016)\citenamefont
  {{Tursunov}}, \citenamefont {{Stuchl{\'\i}k}},\ and\ \citenamefont
  {{Kolo{\v{s}}}}}]{Tursunov:PRD:2016}%
  \BibitemOpen
  \bibfield  {author} {\bibinfo {author} {\bibfnamefont {A.}~\bibnamefont
  {{Tursunov}}}, \bibinfo {author} {\bibfnamefont {Z.}~\bibnamefont
  {{Stuchl{\'\i}k}}}, \ and\ \bibinfo {author} {\bibfnamefont {M.}~\bibnamefont
  {{Kolo{\v{s}}}}},\ }\href {\doibase 10.1103/PhysRevD.93.084012} {\bibfield
  {journal} {\bibinfo  {journal} {Phys. Rev. D}\ }\textbf {\bibinfo {volume}
  {93}},\ \bibinfo {eid} {084012} (\bibinfo {year} {2016})},\ \Eprint
  {http://arxiv.org/abs/1603.07264} {arXiv:1603.07264 [gr-qc]} \BibitemShut
  {NoStop}%
\bibitem [{\citenamefont {{T{\"o}r{\"o}k}}\ and\ \citenamefont
  {{Stuchl{\'\i}k}}(2005)}]{Torok:AA:2005}%
  \BibitemOpen
  \bibfield  {author} {\bibinfo {author} {\bibfnamefont {G.}~\bibnamefont
  {{T{\"o}r{\"o}k}}}\ and\ \bibinfo {author} {\bibfnamefont {Z.}~\bibnamefont
  {{Stuchl{\'\i}k}}},\ }\href {\doibase 10.1051/0004-6361:20052825} {\bibfield
  {journal} {\bibinfo  {journal} {Astronomy and Astrophysics}\ }\textbf
  {\bibinfo {volume} {437}},\ \bibinfo {pages} {775} (\bibinfo {year}
  {2005})},\ \Eprint {http://arxiv.org/abs/astro-ph/0502127}
  {arXiv:astro-ph/0502127 [astro-ph]} \BibitemShut {NoStop}%
\bibitem [{\citenamefont {{T{\"o}r{\"o}k}}\ \emph {et~al.}(2005)\citenamefont
  {{T{\"o}r{\"o}k}}, \citenamefont {{Abramowicz}}, \citenamefont
  {{Klu{\'z}niak}},\ and\ \citenamefont {{Stuchl{\'\i}k}}}]{Torok:AA:2005a}%
  \BibitemOpen
  \bibfield  {author} {\bibinfo {author} {\bibfnamefont {G.}~\bibnamefont
  {{T{\"o}r{\"o}k}}}, \bibinfo {author} {\bibfnamefont {M.~A.}\ \bibnamefont
  {{Abramowicz}}}, \bibinfo {author} {\bibfnamefont {W.}~\bibnamefont
  {{Klu{\'z}niak}}}, \ and\ \bibinfo {author} {\bibfnamefont {Z.}~\bibnamefont
  {{Stuchl{\'\i}k}}},\ }\href {\doibase 10.1051/0004-6361:20047115} {\bibfield
  {journal} {\bibinfo  {journal} {Astronomy and Astrophysics}\ }\textbf
  {\bibinfo {volume} {436}},\ \bibinfo {pages} {1} (\bibinfo {year}
  {2005})}\BibitemShut {NoStop}%
\bibitem [{\citenamefont {{Bambi}}(2012)}]{Bambi:JCAP:2012}%
  \BibitemOpen
  \bibfield  {author} {\bibinfo {author} {\bibfnamefont {C.}~\bibnamefont
  {{Bambi}}},\ }\href {\doibase 10.1088/1475-7516/2012/09/014} {\bibfield
  {journal} {\bibinfo  {journal} {J. Cosmol. Astropart. Phys.}\ }\textbf
  {\bibinfo {volume} {9}},\ \bibinfo {eid} {014} (\bibinfo {year} {2012})},\
  \Eprint {http://arxiv.org/abs/1205.6348} {arXiv:1205.6348 [gr-qc]}
  \BibitemShut {NoStop}%
\bibitem [{\citenamefont {{Bambi}}\ and\ \citenamefont
  {{Nampalliwar}}(2016)}]{Bambi:EPL:2016}%
  \BibitemOpen
  \bibfield  {author} {\bibinfo {author} {\bibfnamefont {C.}~\bibnamefont
  {{Bambi}}}\ and\ \bibinfo {author} {\bibfnamefont {S.}~\bibnamefont
  {{Nampalliwar}}},\ }\href {\doibase 10.1209/0295-5075/116/30006} {\bibfield
  {journal} {\bibinfo  {journal} {Europhys. Lett.}\ }\textbf {\bibinfo {volume}
  {116}},\ \bibinfo {pages} {30006} (\bibinfo {year} {2016})},\ \Eprint
  {http://arxiv.org/abs/1604.02643} {arXiv:1604.02643 [gr-qc]} \BibitemShut
  {NoStop}%
\bibitem [{\citenamefont {{Stuchl{\'{\i}}k}}\ and\ \citenamefont {{Kolo{\v
  s}}}(2015)}]{Stuchlik:MNRAS:2015}%
  \BibitemOpen
  \bibfield  {author} {\bibinfo {author} {\bibfnamefont {Z.}~\bibnamefont
  {{Stuchl{\'{\i}}k}}}\ and\ \bibinfo {author} {\bibfnamefont {M.}~\bibnamefont
  {{Kolo{\v s}}}},\ }\href {\doibase 10.1093/mnras/stv1120} {\bibfield
  {journal} {\bibinfo  {journal} {MNRAS}\ }\textbf {\bibinfo {volume} {451}},\
  \bibinfo {pages} {2575} (\bibinfo {year} {2015})},\ \Eprint
  {http://arxiv.org/abs/1603.07339} {arXiv:1603.07339 [astro-ph.HE]}
  \BibitemShut {NoStop}%
\bibitem [{\citenamefont {{Stuchl{\'\i}k}}\ and\ \citenamefont
  {{Kotrlov{\'a}}}(2009)}]{Stuchlik:GRG:2009}%
  \BibitemOpen
  \bibfield  {author} {\bibinfo {author} {\bibfnamefont {Z.}~\bibnamefont
  {{Stuchl{\'\i}k}}}\ and\ \bibinfo {author} {\bibfnamefont {A.}~\bibnamefont
  {{Kotrlov{\'a}}}},\ }\href {\doibase 10.1007/s10714-008-0709-2} {\bibfield
  {journal} {\bibinfo  {journal} {Gen. Relativ. Gravitation}\ }\textbf
  {\bibinfo {volume} {41}},\ \bibinfo {pages} {1305} (\bibinfo {year}
  {2009})},\ \Eprint {http://arxiv.org/abs/0812.5066} {arXiv:0812.5066
  [astro-ph]} \BibitemShut {NoStop}%
\bibitem [{\citenamefont {{Stuchl{\'\i}k}}\ and\ \citenamefont
  {{Kolo{\v{s}}}}(2016)}]{Stuchlik:APJ:2016}%
  \BibitemOpen
  \bibfield  {author} {\bibinfo {author} {\bibfnamefont {Z.}~\bibnamefont
  {{Stuchl{\'\i}k}}}\ and\ \bibinfo {author} {\bibfnamefont {M.}~\bibnamefont
  {{Kolo{\v{s}}}}},\ }\href {\doibase 10.3847/0004-637X/825/1/13} {\bibfield
  {journal} {\bibinfo  {journal} {Astrophys. J.}\ }\textbf {\bibinfo {volume}
  {825}},\ \bibinfo {eid} {13} (\bibinfo {year} {2016})},\ \Eprint
  {http://arxiv.org/abs/1608.01659} {arXiv:1608.01659 [astro-ph.HE]}
  \BibitemShut {NoStop}%
\bibitem [{\citenamefont {{Stuchl{\'{\i}}k}}\ and\ \citenamefont {{Kolo{\v
  s}}}(2016)}]{Stuchlik:AA:2016}%
  \BibitemOpen
  \bibfield  {author} {\bibinfo {author} {\bibfnamefont {Z.}~\bibnamefont
  {{Stuchl{\'{\i}}k}}}\ and\ \bibinfo {author} {\bibfnamefont {M.}~\bibnamefont
  {{Kolo{\v s}}}},\ }\href {\doibase 10.1051/0004-6361/201526095} {\bibfield
  {journal} {\bibinfo  {journal} {Astronomy and Astrophysics}\ }\textbf
  {\bibinfo {volume} {586}},\ \bibinfo {eid} {A130} (\bibinfo {year} {2016})},\
  \Eprint {http://arxiv.org/abs/1603.07366} {arXiv:1603.07366 [astro-ph.HE]}
  \BibitemShut {NoStop}%
\bibitem [{\citenamefont {{Erez}}\ and\ \citenamefont {{Rosen}}(1959)}]{ER}%
  \BibitemOpen
  \bibfield  {author} {\bibinfo {author} {\bibfnamefont {G.}~\bibnamefont
  {{Erez}}}\ and\ \bibinfo {author} {\bibfnamefont {N.}~\bibnamefont
  {{Rosen}}},\ }\href@noop {} {\bibfield  {journal} {\bibinfo  {journal} {Bull.
  Res. Council Israel}\ }\textbf {\bibinfo {volume} {8F}},\ \bibinfo {pages}
  {47} (\bibinfo {year} {1959})}\BibitemShut {NoStop}%
\bibitem [{\citenamefont {{Bardeen}}\ \emph {et~al.}(1972)\citenamefont
  {{Bardeen}}, \citenamefont {{Press}},\ and\ \citenamefont
  {{Teukolsky}}}]{Bardeen:APJ:1972}%
  \BibitemOpen
  \bibfield  {author} {\bibinfo {author} {\bibfnamefont {J.~M.}\ \bibnamefont
  {{Bardeen}}}, \bibinfo {author} {\bibfnamefont {W.~H.}\ \bibnamefont
  {{Press}}}, \ and\ \bibinfo {author} {\bibfnamefont {S.~A.}\ \bibnamefont
  {{Teukolsky}}},\ }\href {\doibase 10.1086/151796} {\bibfield  {journal}
  {\bibinfo  {journal} {Astrophysical Journal}\ }\textbf {\bibinfo {volume}
  {178}},\ \bibinfo {pages} {347} (\bibinfo {year} {1972})}\BibitemShut
  {NoStop}%
\bibitem [{\citenamefont {{Hod}}(2013)}]{Hod:PRD:2013}%
  \BibitemOpen
  \bibfield  {author} {\bibinfo {author} {\bibfnamefont {S.}~\bibnamefont
  {{Hod}}},\ }\href {\doibase 10.1103/PhysRevD.88.087502} {\bibfield  {journal}
  {\bibinfo  {journal} {Phys. Rev. D}\ }\textbf {\bibinfo {volume} {88}},\
  \bibinfo {eid} {087502} (\bibinfo {year} {2013})}\BibitemShut {NoStop}%
\bibitem [{\citenamefont {{Herrera}}(2005)}]{Herrera:FPL:2005}%
  \BibitemOpen
  \bibfield  {author} {\bibinfo {author} {\bibfnamefont {L.}~\bibnamefont
  {{Herrera}}},\ }\href {\doibase 10.1007/s10702-005-2467-7} {\bibfield
  {journal} {\bibinfo  {journal} {Found. Phys. Lett.}\ }\textbf {\bibinfo
  {volume} {18}},\ \bibinfo {pages} {21} (\bibinfo {year} {2005})},\ \Eprint
  {http://arxiv.org/abs/gr-qc/0402052} {gr-qc/0402052} \BibitemShut {NoStop}%
\bibitem [{\citenamefont {{Toshmatov}}\ \emph {et~al.}(2017)\citenamefont
  {{Toshmatov}}, \citenamefont {{Stuchl{\'\i}k}},\ and\ \citenamefont
  {{Ahmedov}}}]{Toshmatov:PRD:2017}%
  \BibitemOpen
  \bibfield  {author} {\bibinfo {author} {\bibfnamefont {B.}~\bibnamefont
  {{Toshmatov}}}, \bibinfo {author} {\bibfnamefont {Z.}~\bibnamefont
  {{Stuchl{\'\i}k}}}, \ and\ \bibinfo {author} {\bibfnamefont {B.}~\bibnamefont
  {{Ahmedov}}},\ }\href {\doibase 10.1103/PhysRevD.95.084037} {\bibfield
  {journal} {\bibinfo  {journal} {Phys. Rev. D}\ }\textbf {\bibinfo {volume}
  {95}},\ \bibinfo {eid} {084037} (\bibinfo {year} {2017})},\ \Eprint
  {http://arxiv.org/abs/1704.07300} {arXiv:1704.07300 [gr-qc]} \BibitemShut
  {NoStop}%
\bibitem [{\citenamefont {{Boshkayev}}\ \emph {et~al.}(2018)\citenamefont
  {{Boshkayev}}, \citenamefont {{Muccino}}, \citenamefont {{Rueda}},\ and\
  \citenamefont {{Zhumakhanova}}}]{Boshkayev}%
  \BibitemOpen
  \bibfield  {author} {\bibinfo {author} {\bibfnamefont {K.~A.}\ \bibnamefont
  {{Boshkayev}}}, \bibinfo {author} {\bibfnamefont {M.}~\bibnamefont
  {{Muccino}}}, \bibinfo {author} {\bibfnamefont {J.~A.}\ \bibnamefont
  {{Rueda}}}, \ and\ \bibinfo {author} {\bibfnamefont {G.~D.}\ \bibnamefont
  {{Zhumakhanova}}},\ }\href@noop {} {\bibfield  {journal} {\bibinfo  {journal}
  {arXiv e-prints}\ ,\ \bibinfo {eid} {arXiv:1802.06773}} (\bibinfo {year}
  {2018})},\ \Eprint {http://arxiv.org/abs/1802.06773} {arXiv:1802.06773
  [astro-ph.HE]} \BibitemShut {NoStop}%
\bibitem [{\citenamefont {{Event Horizon Telescope Collaboration}}\ \emph
  {et~al.}(2019)\citenamefont {{Event Horizon Telescope Collaboration}},
  \citenamefont {{Akiyama}},\ and\ \citenamefont {et~al.}}]{EHT}%
  \BibitemOpen
  \bibfield  {author} {\bibinfo {author} {\bibnamefont {{Event Horizon
  Telescope Collaboration}}}, \bibinfo {author} {\bibfnamefont
  {K.}~\bibnamefont {{Akiyama}}}, \ and\ \bibinfo {author} {\bibnamefont
  {et~al.}},\ }\href {\doibase 10.3847/2041-8213/ab0ec7} {\bibfield  {journal}
  {\bibinfo  {journal} {Astrophys. J. Lett.}\ }\textbf {\bibinfo {volume}
  {875}},\ \bibinfo {eid} {L1} (\bibinfo {year} {2019})}\BibitemShut {NoStop}%
\bibitem [{\citenamefont {{Gravity Collaboration}}\ \emph
  {et~al.}(2018)\citenamefont {{Gravity Collaboration}}, \citenamefont
  {{Abuter}},\ and\ \citenamefont {et~al.}}]{AAP}%
  \BibitemOpen
  \bibfield  {author} {\bibinfo {author} {\bibnamefont {{Gravity
  Collaboration}}}, \bibinfo {author} {\bibfnamefont {R.}~\bibnamefont
  {{Abuter}}}, \ and\ \bibinfo {author} {\bibnamefont {et~al.}},\ }\href
  {\doibase 10.1051/0004-6361/201833718} {\bibfield  {journal} {\bibinfo
  {journal} {Astronomy \& Astrophysics}\ }\textbf {\bibinfo {volume} {615}},\
  \bibinfo {eid} {L15} (\bibinfo {year} {2018})},\ \Eprint
  {http://arxiv.org/abs/1807.09409} {arXiv:1807.09409 [astro-ph.GA]}
  \BibitemShut {NoStop}%
\end{thebibliography}%

\end{document}